\documentclass[conference]{IEEEtran}

\IEEEoverridecommandlockouts
\usepackage[nocompress]{cite}

\usepackage[hidelinks]{hyperref}
\usepackage{amsmath,amssymb,amsfonts,bm}
\usepackage{algpseudocode}
\usepackage{algorithm}
\usepackage{graphicx}
\usepackage{subfigure}
\usepackage{textcomp}
\usepackage{xcolor}
\usepackage{comment}
\usepackage{amsthm}
\usepackage{multicol,xparse,environ}
\usepackage{enumitem}
\usepackage{ifthen}

\NewEnviron{auxmulticols}[1]{%
  \ifnum#1<2\relax
    \BODY
  \else
    \begin{multicols}{#1}
      \BODY
    \end{multicols}%
  \fi
}

\def\BibTeX{{\rm B\kern-.05em{\sc i\kern-.025em b}\kern-.08em
    T\kern-.1667em\lower.7ex\hbox{E}\kern-.125emX}}

\newtheorem{theorem}{Theorem}
\newtheorem{lemma}{Lemma}
\theoremstyle{plain}

\newcommand{\fullpaperurl}{http://arxiv.org/abs/2009.05766}
\newcommand*{\fullpaperterm}{\href{\fullpaperurl}{full paper}}

\newcommand*{\ourtech}{NetMax}
\newcommand*{\prague}{Prague}
\newcommand*{\allreduce}{Allreduce-SGD}
\newcommand*{\adpsgd}{AD-PSGD}
\newcommand*{\coordinator}{Network Monitor}
\newcommand*{\resnet}{ResNet18}
\newcommand*{\vgg}{VGG19}
\newcommand*{\cifar}{CIFAR10}

\newboolean{isfullpaper}

\begin{document}
\setboolean{isfullpaper}{true}

\ifthenelse{\boolean{isfullpaper}}
{}
{
}

\title{Communication-efficient Decentralized Machine Learning over Heterogeneous Networks}

\author{\IEEEauthorblockN{Pan Zhou\textsuperscript{1}\textsuperscript{*}\thanks{*This work was done when this author was a visiting student at National University of Singapore.}, Qian Lin\textsuperscript{2}, Dumitrel Loghin\textsuperscript{2}, Beng Chin Ooi\textsuperscript{2}, Yuncheng Wu\textsuperscript{2}, Hongfang Yu\textsuperscript{1}}
	\IEEEauthorblockA{\textsuperscript{1} University of Electronic Science and Technology of China\\}
	
	\IEEEauthorblockA{\textsuperscript{2} National University of Singapore\\}
	
	
	\IEEEauthorblockA{\textit{willzhoupan@gmail.com, \{linqian, dumitrel, ooibc, wuyc\}@comp.nus.edu.sg, yuhf@uestc.edu.cn} \\}
}


\maketitle

\begin{abstract}
In the last few years, distributed machine learning has been usually executed over heterogeneous networks such as a local area network within a multi-tenant cluster or a wide area network connecting data centers and edge clusters. In these heterogeneous networks, the link speeds among worker nodes vary significantly, making it challenging for state-of-the-art machine learning approaches to perform efficient training. Both centralized and decentralized training approaches suffer from low-speed links. In this paper, we propose a decentralized approach, namely \ourtech{}, that enables worker nodes to communicate via high-speed links and, thus, significantly speed up the training process. \ourtech{} possesses the following novel features. First, it consists of a novel consensus algorithm that allows worker nodes to train model copies on their local dataset asynchronously and exchange information via peer-to-peer communication to synchronize their local copies, instead of a central master node (i.e., parameter server). Second, each worker node selects one peer randomly with a fine-tuned probability to exchange information per iteration. In particular, peers with high-speed links are selected with high probability. Third, the probabilities of selecting peers are designed to minimize the total convergence time. Moreover, we mathematically prove the convergence of \ourtech{}. We evaluate \ourtech{} on heterogeneous cluster networks and show that it achieves speedups of 3.7$\times$, 3.4$\times$, and 1.9$\times$ in comparison with the state-of-the-art decentralized training approaches Prague, Allreduce-SGD, and AD-PSGD, respectively.
\end{abstract}


\setcounter{page}{1}
\setcounter{figure}{0}

\section{Introduction}\label{sec:introduction}
Recently, distributed machine learning has become increasingly popular.
%
The training phase of distributed machine learning is typically executed in a multi-tenant cluster~\cite{xiao2018gandiva, jeon2019analysis, OoiTWWCCGLTWXZZ15, CaiGZACOTTW18} or across data centers and edge clusters~\cite{hsieh2017gaia,Zheng2020TRACERAF}. 
As a result, distributed machine learning faces the emerging problem of communication over heterogeneous networks, where the link speeds among worker nodes are highly different.
For machine learning model training in a multi-tenant cluster, one of the main challenges is avoiding resource fragmentation by allocating adequate co-located GPUs for worker nodes~\cite{jeon2019analysis,Wang2016DatabaseMD}.
Often, the allocated GPUs are placed across machines or even racks~\cite{gu2019tiresias}. 
The link speeds within a machine differ significantly from those across machines or racks due to different capacities of the underlying infrastructure. 
Moreover, the network contention among distributed learning jobs can easily cause network congestion even on a 100 Gbps network, making the difference in link speeds more significant~\cite{gu2019tiresias, xiao2018gandiva}. 
For model training across data centers or edge clusters, the link speeds of wide area network (WAN) connecting the worker nodes are mainly determined by the geographic distance. 
For example, the link speed between geographically-close data centers could be up to 12$\times$ faster than between distant ones~\cite{zhou2019privacy}. 
In such highly heterogeneous networks, fully utilizing high-speed links is important to reduce the communication cost and thus accelerate the distributed training.

\begin{figure}[!t]
	\centering
	\includegraphics[width=0.46\textwidth]{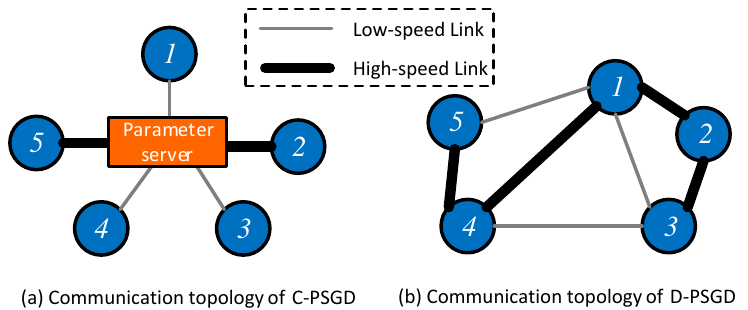}
	\caption{Communication topology of C-PSGD and D-PSGD. For C-PSGD, worker nodes communicate with a central parameter server to synchronize their training. For D-PSGD, worker nodes communicate with each other in a graph. Both C-PSGD and D-PSGD might be limited by low-speed links.}
	\label{fig:example-bottleneck} 
	\vspace{-1ex}
\end{figure}

\begin{figure}[!t]
	\centering
	\includegraphics[width=0.48\textwidth]{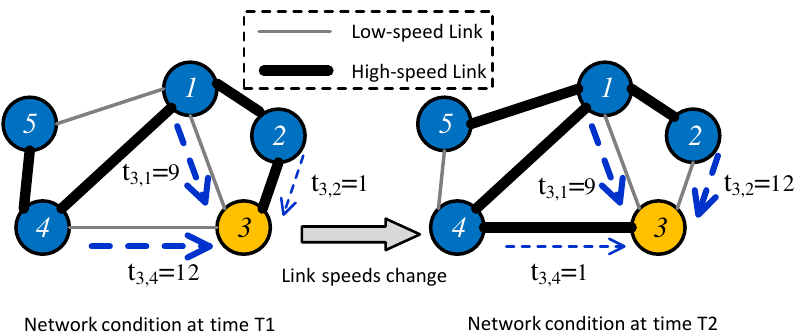}
	\caption{Heterogeneous and dynamic network condition example.}
	\label{fig:example-bottleneck-2} 
	\vspace{-1ex}
\end{figure}

Traditional centralized parallel stochastic gradient descent (C-PSGD) approaches, such as Adam \cite{chilimbi2014project} and R2SP~\cite{chen2019round}, require a central node called parameter server (PS) to maintain a global model.
At each iteration, all the worker nodes exchange information (i.e., parameters and gradients) with the PS to update the global model, as shown in \autoref{fig:example-bottleneck}(a).
%
%
The training speed of these approaches is typically limited by (1) the network congestion at the central PS, and (2) the low-speed links between the worker nodes and PS since a worker node needs to wait for the updated model from the PS before proceeding to the next iteration.
Existing decentralized parallel stochastic gradient descent (D-PSGD) approaches~\cite{jia2018highly,tang2018d,xin2019decentralized,lian2018asynchronous,hegedHus2019gossip,lalitha2019peer} avoid the bottleneck of PS by adopting peer-to-peer communication in a connected graph, as shown in \autoref{fig:example-bottleneck}(b). 
However, they may still suffer from the low-speed links.
In synchronous D-PSGD approaches, such as Allreduce-SGD~\cite{jia2018highly} and D2~\cite{tang2018d}, the worker nodes communicate with their neighbors to exchange information at each iteration, but all the worker nodes need to enter the next iteration simultaneously. Thus, the worker nodes with low-speed links will restrict the training process.
%
%
%
%
%
In asynchronous D-PSGD approaches, such as AD-PSGD~\cite{lian2018asynchronous}, GoSGD~\cite{hegedHus2019gossip}, and Prague~\cite{luo2020prague}, a worker node randomly chooses one or several neighbors to communicate with, and starts the next iteration once these communications are completed, independently from the executions of other worker nodes.  
Nevertheless, the worker node may frequently communicate over low-speed links, resulting in a high communication cost. 
Take the left side of \autoref{fig:example-bottleneck-2} as an example, where $t_{i,j}$ denotes the network latency from node $j$ to node $i$. 
If node $3$ chooses neighbor nodes uniformly random, then in about $67\%$ of its iterations, it pulls models from nodes $1$ and $4$, but this is time-consuming due to the high network latency.
%
Recently, SAPS-PSGD~\cite{tang2020communication} improves AD-PSGD and GoSGD by letting the worker nodes communicate with each other based on a fixed network subgraph that consists of initially high-speed links.
However, the link speeds may dynamically change according to the arrival/leaving of other learning jobs or the network bandwidth reallocation.
As a consequence, a link evaluated to be of high-speed at the start of the targeted training may become low-speed at another time.
As shown in \autoref{fig:example-bottleneck-2}, if node $3$ uses the subgraph of initially high-speed links observed at time T1, then it will only communicate with node $2$.
However, this link may become low-speed at time T2, resulting in a high communication cost.

In this paper, we aim to answer the following research question: \emph{is it possible to fully use the high-speed links in dynamic heterogeneous networks to accelerate distributed training while providing convergence guarantees in theory?}
To tackle this problem, we design a novel communication-efficient decentralized asynchronous distributed machine learning approach (\ourtech{}) consisting of a consensus SGD algorithm and a communication policy generation algorithm.
The \ourtech{} approach is also a part of our federated learning \cite{WuCXCO20} system \textit{Falcon}\footnote{\label{note:falcon}\url{https://www.comp.nus.edu.sg/~dbsystem/fintech-Falcon/}}, to improve the communication efficiency during the decentralized training process across data centers located in multiple organizations. 
\ourtech{} enables worker nodes to communicate preferably through high-speed links in a dynamic heterogeneous network to accelerate the training. 
In addition to providing guarantees of convergence for the distributed training, \ourtech{} also has the following three novelties.
(1)~Worker nodes iteratively and asynchronously train model copies on their local data by using the consensus SGD algorithm. Each worker node randomly selects only one neighbor to communicate with at each iteration.
(2)~The neighbors with high-speed links are selected with high probabilities, and those with low-speed links are selected with low probabilities. Such a design enables the worker nodes not only to communicate preferably through high-speed links but also to detect the network dynamics. 
%
(3)~The communication policy generation algorithm updates the communication probabilities for selecting neighbors according to the detected link speeds, which can make the training adapt to network dynamics. Importantly, we analyze the relationship between the convergence of consensus SGD and the communication probabilities. The communication policy generation algorithm is designed such that the total convergence time can be minimized.

The main contributions of this paper are as follows.

\begin{itemize}
	
	\item We identify the challenges of distributed learning over a heterogeneous network and propose a communication-efficient asynchronous D-PSGD approach, namely \ourtech{}, to fully utilize fast links and speed up the training. 
	
	\item We theoretically prove that \ourtech{} provides convergence guarantees for decentralized training.
	
	\item We implement \ourtech{} and evaluate it on a heterogeneous multi-tenant cluster. 
	Experimental results show that \ourtech{} achieves up to 3.7$\times$ speedup over state-of-the-art approaches.
\end{itemize}

The rest of the paper is organized as follows. \autoref{sec:preliminaries} formulates our decentralized training problem. \autoref{sec:design} presents the design of \ourtech{}. \autoref{sec:convergence-analysis} provides the theoretical analysis of \ourtech{}. \autoref{sec:eval} evaluates \ourtech{} over both homogeneous and heterogeneous networks. \autoref{sec:relwork} presents related works and \autoref{sec:conclusion} concludes the paper.
\section{Preliminaries}
\label{sec:preliminaries}

\subsection{Problem Formulation}

We first formulate the problem using the notations in \autoref{tab:notations}.
Let $M$ denote the number of worker nodes in the network. 
The communication topology connecting the nodes can be represented as an undirected graph $\mathcal{\boldsymbol{G}}$ with vertex set $\mathcal{\boldsymbol{V}}=\{1,2,\dots,M\}$ and edge set $\mathcal{\boldsymbol{E}}=\mathcal{\boldsymbol{V}}\times\mathcal{\boldsymbol{V}}$. 
Each worker node $i \in [1,M]$ holds a dataset $\mathcal{D}_i$. 
To collaboratively train a model on the joint dataset, each worker node trains a local model using its own dataset and exchanges the model information with its neighbors in the graph. 
We formulate such decentralized training as an unconstrained consensus optimization problem:

\begin{table}[!t]
	\centering
	\caption{Summary of notations}
	\begin{tabular}{l|p{0.38\textwidth}}
		\hline
		Notation & Description\\
		\hline
		$M$&the number of distributed worker nodes\\
		$d_{i,m}$& neighborhood indicator between node $i$ and $m$ \\
		$F$&global optimization objective function\\
		$f$&local loss function of worker nodes\\
		$x_{i}$&local model parameters of node $i$\\
		$\xi_{i}$&local gradient noise of node $i$\\
		$\rho$& weight of model difference between worker nodes\\
		$\alpha$&SGD learning rate\\
		$k$&the global iteration step\\
		$p_{i,m}$& the probability for node $i$ to select node $m$ \\
		$p_{i}$&the probability that node $i$ communicates with one of its neighbors at a global step \\
		$t_{i,m}$&iteration time for node $i$ communicating with node $m$\\
		$\overline{t}_{i}$&local average iteration time at node $i$\\
		$\overline{t}$&the global average iteration time\\
		\hline
	\end{tabular}
	\label{tab:notations}
	\vspace{-1ex}
\end{table}

\begin{equation} 
\min_{x} F(x)=\sum_{i=1}^{M} \Big[ \mathbb{E}[f(x_i;\mathcal
{D}_i)] + \frac{\rho}{4}\sum_{m=1}^{M} d_{i,m} \big\lVert x_i \! -\! x_{m} \big\rVert^{2}\Big] \label{eq:a1_1}
\end{equation}
where $x_i$ and $f(x_i;\mathcal{D}_i)$ are the model parameters and the loss function of worker node $i$, respectively. $d_{i,m}$ is a connection indicator such that $d_{i,m} = 1$ if the worker nodes $i$ and $m$ are neighbors, and $d_{i,m} = 0$ otherwise. $\lVert x_i - x_{m} \rVert^{2}$ measures the model difference between worker node $i$ and $m$, if they are neighbors. The weight $\rho$ indicates the importance of model difference between neighbors in the objective function. The goal of Eq.~(\ref{eq:a1_1}) is to minimize the loss functions of distributed worker nodes as well as their model differences, ensuring that all nodes will converge to the same optima.

\subsection{Definitions}\label{subsec:definitions}

\noindent\textbf{Iteration step}. 
Let $n$ be a local iteration step of a worker node, which increases by one when the node communicates with a neighbor. 
Let $k$ be a global iteration step, which advances if a local step of any worker node increases. 
Note that there is only one worker node communicating with a neighbor at any global iteration step.

\begin{figure}[!t]
	\centering
	\includegraphics[width=0.46\textwidth]{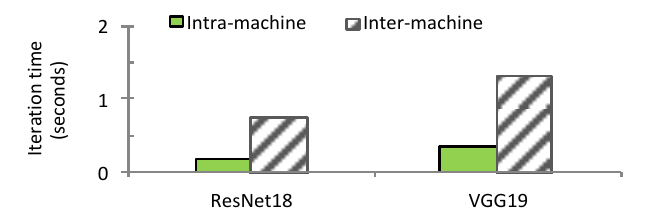}
	\caption{Average iteration time corresponding to intra-machine (fast) and inter-machine (slow) communication. Nodes are connected by 1000 Mbps Ethernet.}
	\label{fig:iteration-time-example} 
	\vspace{-1ex}
\end{figure}

\noindent\textbf{Iteration time.} At any local iteration step of worker node $i$, let $C_i$ be the local computation time, and $N_{i,m}$ be the network communication time between node $i$ and a neighboring node $m$. 
We denote the iteration time (i.e., the total duration time of this local iteration step) as $t_{i,m}$. 
In this paper, we parallelize the local computation and network communication in each worker node, and thus the iteration time is calculated as 
\[
t_{i,m} = \max\{C_{i}, N_{i,m}\}.
\]
We observe that the communication time usually dominates. 
In our experimental setup, detailed in \autoref{sec:eval}, the iteration time corresponding to inter-machine communication (i.e., via slow links) can be up to $4\times$ larger compared to the iteration time corresponding to intra-machine communication (i.e., via fast links). 
\autoref{fig:iteration-time-example} demonstrates such comparison measured for the training of two deep learning models, namely \resnet{} and \vgg{}.
Therefore, network communication through a fast link can result in reduced iteration time. 
%

\noindent
\textbf{Probability.} 
To better utilize high-speed links, we design an adaptive communication strategy that allows a worker node to choose its neighbors with different probabilities, presuming neighbors with high-speed links tend to be selected. 
The probability distribution may change over time with respect to the dynamic network conditions. 
This is in contrast to existing decentralized training algorithms such as GoSGD~\cite{hegedHus2019gossip} and AD-PSGD~\cite{lian2018asynchronous}, which use fixed uniform probability distributions.

Let $\boldsymbol{P}=[p_{i,m}]_{M \times M}$ be the communication policy matrix (i.e., communication probabilities for worker nodes to select their neighbors), where an entry $p_{i,m}$ denotes the communication probability for node $i$ to select node $m$ as its neighbor. 
The $i$-th row in $\boldsymbol{P}$ is the communication probability distribution of node $i$. 
Let $\overline{t}_{i}$ denote the average iteration time for node $i$,
\begin{equation}
\overline{t}_{i} = \sum_{m=1}^{M} t_{i,m} \cdot p_{i,m} \cdot d_{i,m}\label{eq:d3}
\end{equation}

Let $p_{i}$ denote the probability of worker node $i$ communicating with one of its neighbors at any global step, which can be derived as
\begin{equation}
p_{i} =\frac{1/\overline{t}_{i}}{\sum_{m=1}^{M}(1/\overline{t}_{m})}  \label{eq:d4}
\end{equation}
where $1/\overline{t}_{i}$ is the iterative frequency of worker node $i$. The larger iterative frequency a worker node has, the higher probability (\textit{w.r.t.} $p_i$) it can communicate at a global iteration step.
According to Eq.~(\ref{eq:d3}) and Eq.~(\ref{eq:d4}), $p_i$ is a function of $p_{i,m}$. Thus, the design of the communication policy affects the communication frequencies of the worker nodes. 

\section{Approach}
\label{sec:design}

In this section, we present our communication-efficient asynchronous decentralized machine learning approach, termed \ourtech{}. 

\subsection{Overview}\label{subsec:overview}

\begin{figure}[!t]
	\centering
	\includegraphics[width=0.49\textwidth]{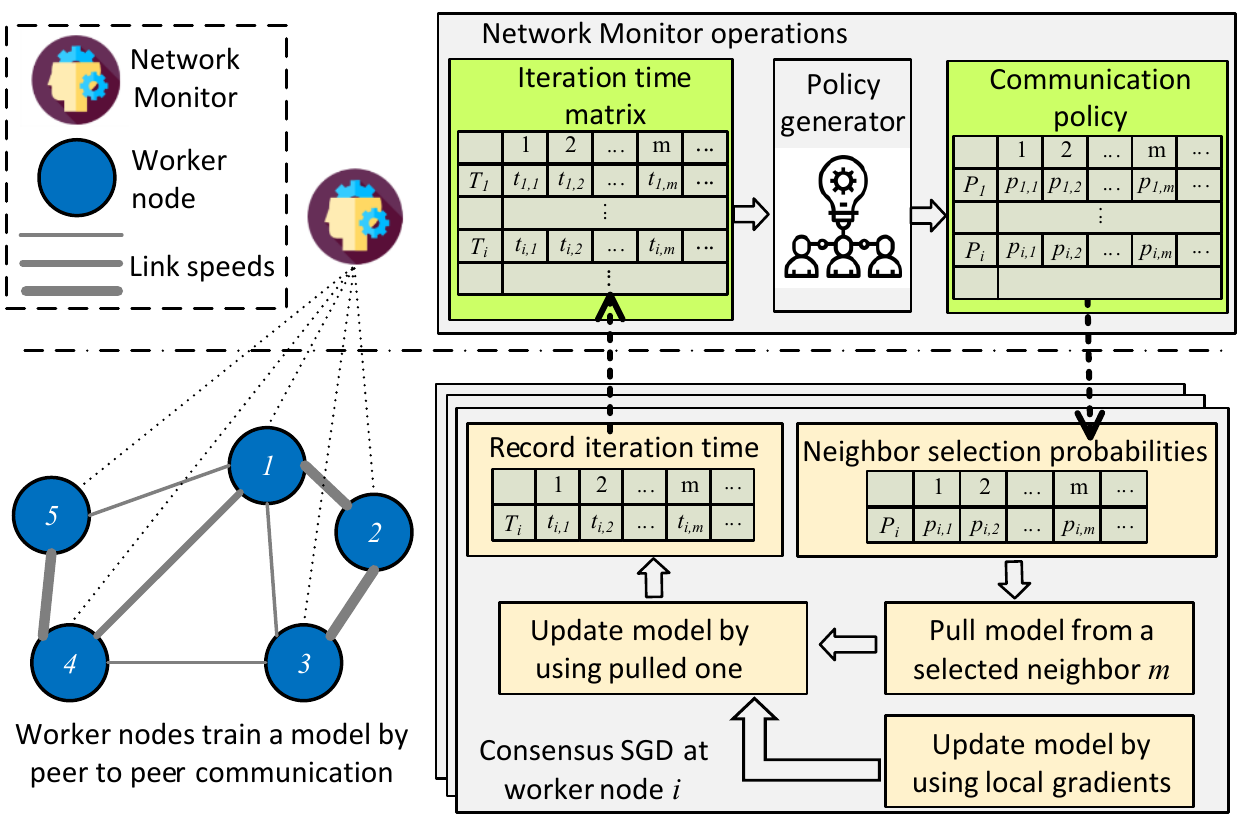}
	\caption{Overview of the \ourtech{} design.}
	\label{fig:overview} 
	\vspace{-1ex}
\end{figure}

\autoref{fig:overview} illustrates the overall design of \ourtech{}. 
The worker nodes collaboratively train a model on their local data in a decentralized manner. 
To adapt to the dynamic network environment, worker nodes rely on a central \coordinator{} to track the network status so that they can optimize the utilization of high-speed links to accelerate the training process. 
One noteworthy aspect is that the central \coordinator{} of \ourtech{} does not collect any worker node's training data or model parameters; instead, it only collects a small amount of time-related statistics for evaluating the network condition.
Therefore, \ourtech{} is different from existing centralized training methods (e.g., Adam~\cite{chilimbi2014project} and R2SP~\cite{chen2019round}), where the central server is responsible for synchronizing the worker nodes and aggregating model parameters from them. 
This is much likely to become a bottleneck, as discussed in \autoref{sec:relwork}. 

\noindent\textbf{\coordinator{}.} 
In order to estimate the network condition, the \coordinator{} periodically collects the iteration time of each worker node, as this information reflects the link speed between worker nodes (see \autoref{subsec:definitions}). 
Let $T_s$ be the collection period, which can be adjusted according to the network conditions. 
For example, if the link speeds change quickly, $T_s$ can be set to a small value to react to the network changes.
\autoref{alg:coordinator} presents the computations done by the \coordinator{}. 
In every period, it collects the time statistics from the worker nodes ({lines \ref{code:c3}-\ref{code:c4}}),
and computes a communication policy matrix $\boldsymbol{P}$ and a weight $\rho$ based on the statistics ({line \ref{code:c5}}). 
The entries of the matrix $\boldsymbol{P}$ are the probabilities for worker nodes to select neighbors, while the weight $\rho$ defines the importance of the model difference between worker nodes (see Eq.~\ref{eq:a1_1}).
The policy $\boldsymbol{P}$ and weight $\rho$ are adaptively tuned to make the whole training process converge fast.
The details of policy design will be presented in \autoref{subsection:plicy-design}.
Finally, the \coordinator{} sends the policy matrix $\boldsymbol{P}$ and weight $\rho$ back to the worker nodes (line \ref{code:c6}). 
The \coordinator{} uses the iteration time $t_{i,m}$ to evaluate the link speed between worker nodes $i$ and $m$, rather than measuring the exact speed. The rationale for doing so is twofold. First, the iteration time reflects well the corresponding link speed. A short iteration time means short communication time as well as fast link speed. Second, the iteration time can be directly measured by the worker nodes during their training, which avoids injecting a large amount of traffic into the network to measure the exact link speed. 

\begin{algorithm}[t] 
	\caption{\ourtech{} at \coordinator{}} 
	\label{alg:coordinator} 
	\begin{algorithmic}[1] 
		\begin{small}
			\Require schedule period $T_s$, learning rate $\alpha$, out-loop search round $K$, inner-loop search round $R$.
			\State Initialize iteration time matrix $[t_{i,m}]_{M\times M} \gets [0]_{M\times M}$\label{code:c1}
			\While{TRUE} \label{code:c2}
			\State  Send iteration-time request to worker nodes\label{code:c3}
			\State  $[t_{i,m}]_{M\times M}\gets$ Receive the iteration time \label{code:c4}
			\State $\boldsymbol{P}$, $\rho\gets$ \Call{GeneratePolicyMatrix} {$\alpha,K,R,[t_{i,m}]_{M\times M}$} \label{code:c5}
			\State Send $\boldsymbol{P}$, $\rho$ to worker nodes \label{code:c6}
			\State $Sleep(T_s)$\label{code:c7}
			\EndWhile\\
		\end{small}
	\end{algorithmic}
	\vspace{-1ex}
\end{algorithm} 

\noindent\textbf{Worker Nodes.} 
Each worker node $i$ conducts a consensus SGD algorithm to train a model replica on its local data, as depicted in \autoref{alg:nadsgd}. In each iteration, a worker node updates its local model in two steps. First, it updates the model based on its computed gradients. Second, it selects one of its neighbors (e.g., node $m$) according to the communication policy and pulls model parameters from the neighbor to update its local model again. Meanwhile, it maintains an average iteration time $t_{i,m}$ for the communication with neighbor $m$. When receiving a request from the \coordinator{}, the iteration time is sent for computing the new communication policy. After receiving the new policy from the \coordinator{}, the worker updates its local policy accordingly.

\subsection{The Consensus SGD Algorithm}\label{subsection:psgd-worker}
As our decentralized training optimization problem formulated by Eq.~(\ref{eq:a1_1}) is unconstrained, we can solve it by the standard stochastic gradient descent (SGD) method, where the gradients at worker node $i$ can be derived as
\begin{equation}
\nabla F(x_{i};\mathcal{D}_i) = \nabla f(x_{i};\mathcal{D}_i) +\frac{\rho}{2}\sum_{m=1}^{M} (d_{i,m}\!+\!d_{m,i})(x_{i}-x_{m}) \label{eq:a4}
\end{equation}
Then the parameters at worker node $i$ can be updated by
\begin{equation}
x_{i}^{n+1} = x_{i}^{n} - \alpha \nabla F(x_{i}^{n};\mathcal{D}_{i,n}) \label{eq:a5}
\end{equation}
where $x_{i}^{n}$ are the model parameters and $\mathcal{D}_{i,n}$ is the sampled dataset from $\mathcal{D}_{i}$ of worker node $i$ at local iteration step $n$. Based on Eq.~(\ref{eq:a4}) and (\ref{eq:a5}), the exchanged information between neighbors consists of their model parameters. However, simply applying the update rule in Eq. (\ref{eq:a5}) requires a worker node to pull parameters from all its neighbors, which means that the training process might be limited by the low-speed links. 

To reduce communication time, we propose a consensus SGD algorithm. At each iteration, a worker node only pulls model parameters from a randomly selected neighbor. The neighbors with high-speed links are more likely to be selected. 
As training proceeds, each worker node's local information will be propagated in the communication graph and gradually reach other worker nodes.
Meanwhile, each worker node uses both local gradients and the model from its neighbor to update its model. Therefore, both the loss functions of worker nodes and the model differences between them can be minimized in the training process.
%
Consequently, the worker nodes converge to the same optima.

The detailed executions of a worker node are presented in \autoref{alg:nadsgd}. At the beginning of each iteration, a worker node $i$ first updates its communication probabilities if a new policy is received from the \coordinator{} ({lines \ref{code:l4}-\ref{code:l6}}). Then it randomly selects a neighbor $m$ with probability $p_{i,m}$ and requests the freshest parameters from it ({lines \ref{code:l7}-\ref{code:l8}}). A neighbor with a high-speed link has a higher probability of being selected. Therefore, the overall communication time can be reduced. 

After sending the request, worker node $i$ continues to compute gradients using its local dataset and updates its model using these gradients ({line \ref{code:l9}}). Meanwhile, it keeps checking whether the requested parameters are returned ({line \ref{code:l10}}). When receiving the response from neighbor $m$, it updates the model again using the received parameters ({lines \ref{code:l11}-\ref{code:l13}}). This two-step update overlaps the gradient computation and parameter communication, which further reduces the training time.~
Moreover, if a neighbor is chosen with a lower probability, the model update rule can assign a higher weight to the pulled model from that neighbor, which enables the nodes to maintain enough information from neighbors chosen with low probabilities.

\color{black}
At the end of the iteration, worker node $i$ updates the average iteration time corresponding to the communication with neighbor $m$ ({line\ref{code:l14}}). Specifically, we adopt the exponential moving average (EMA) \cite{Hunter1986} to maintain the iteration time ({lines \ref{code:l17}-\ref{code:l20}}). This is because (1) we do not need to store historical time data, and (2) the window length of moving average can be tuned by changing the smoothing factor $\beta$, where a smaller $\beta$ means a shorter window. The value of $\beta$ can be adapted to the network dynamics. When the link speeds change quickly, $\beta$ should be set to a low value to capture the latest speeds. At the same time, the period of policy computing at the \coordinator{}, $T_s$, should be short too. Then the \coordinator{} can quickly update the communication policy for worker nodes, and the entire system adapts to dynamic network conditions.

\subsection{Communication Policy Generation} \label{subsection:plicy-design}

The communication policy is derived from solving an optimization problem based on the convergence analysis of the proposed decentralized training approach. 
The training on the worker nodes can be written in matrix form as
\begin{equation} 
\boldsymbol{x}^{k+1}=\boldsymbol{D}^{k}(\boldsymbol{x}^{k}-\alpha \boldsymbol{g}^{k})  \label{eq:j1}
\end{equation}
where $\boldsymbol{x}^{k}$ and $\boldsymbol{g}^{k}$ are the vectors containing all worker nodes' model parameters and gradients at global step $k$, respectively. $\boldsymbol{D}^{k}$ is a random matrix at global step $k$ related to the communication policy $\boldsymbol{P}$. The models' deviation from the optima $x^*$ can be bounded as follows,
\begin{equation}
\mathbb{E}[ \lVert \bm{x}^{k}-x^{*}\boldsymbol{1}\rVert^{2}] \leq \lambda^{k} \lVert \bm{x}^{0}-x^{*}\boldsymbol{1} \rVert^{2}+ \alpha^{2}\sigma^{2} \frac{\lambda}{1-\lambda} \label{eq:j2}
\end{equation}
where $\boldsymbol{1}$ denotes a vector with entries equal to 1 and $\lambda$ is the second largest eigenvalue of matrix $\mathbb{E}[(\boldsymbol{D}^k)^{T} \boldsymbol{D}^k]$. The detailed derivation of Eq. (\ref{eq:j1}) and (\ref{eq:j2}) is presented in \autoref{sec:convergence-analysis}. 

A noteworthy aspect is that most existing decentralized machine learning algorithms, such as Gossiping SGD~\cite{jin2016scale} and GoSGD~\cite{hegedHus2019gossip}, can be formulated in the form of Eq. (\ref{eq:j1}). Their models' deviation from the optima can also be bounded by Eq. (\ref{eq:j2}). In these algorithms, the worker nodes choose their neighbors uniformly at random, resulting in the fastest convergence rate over a homogeneous network\cite{boyd2006randomized}. In other words, they have small global iteration steps $k$ to convergence. 
However, in a heterogeneous network, the average global step time $\overline{t}$ can be very long since the worker nodes frequently choose neighbors with low-speed links. As a result, the total time to convergence, $k\overline{t}$, is very long. In contrast, choosing high-speed-link neighbors with high probability could reduce the average iteration time $\overline{t}$, but may lead to a slow convergence rate (i.e., larger global iteration steps $k$ to convergence). Therefore, there exists a trade-off between fast convergence rate and short iteration time to minimize the total time $k\overline{t}$ to convergence.

\begin{algorithm}[t] 
	\caption{\ourtech{} at worker node $i$}
	\label{alg:nadsgd} 
	\begin{algorithmic}[1] %
		\begin{small}
			\Require learning rate $\alpha$, weight $\rho$, iteration number $N$, smoothing factor $\beta$.
			\Ensure Trained model $x_{i}$
			\State  Initialize model $x_{i}^{0}$ \label{code:l0}
			\State Initialize probabilities $[p_{i,1},p_{i,2},\dots,p_{i,M}]\gets[1/M]_{1\times M}$ \label{code:l1}
			\State Initialize iteration time vector $\boldsymbol{T}_i\gets[0]_{1\times M}$ \label{code:l2}
			\For{$n \in \{1,2,\dots,N\}$} \label{code:l3}
			\If {received new policy} \label{code:l4}
			\State $[p_{i,1},p_{i,2},\dots,p_{i,M}] \gets recv.\boldsymbol{P}[i]$ \label{code:l5}
			\State $\rho \gets recv.\rho$ \label{code:l5_1}
			\EndIf \label{code:l6}
			\State Random select a neighbor $m$ with probability $p_{i,m}$ \label{code:l7}
			\State Request parameter $x_{m}$ from worker node $m$ \label{code:l8}
			\State $x_{i}^{n} \gets x_{i}^{n}-\alpha\nabla f(x_{i}^{n};\mathcal{D}_{i,n})$ \label{code:l9} \qquad$\triangleright$ First step update
			\State Wait parameter $x_{m}$\label{code:l10}
			\State $\theta_{i}^{n} \gets \frac{\rho}{2}\frac{d_{i,m}+d_{m,i}}{p_{i,m}}(x_{i}^n - x_{m})$\label{code:l11}
			\State $x_{i}^{n} \gets x_{i}^{n} - \alpha\theta_{i}^{n} $\label{code:l12} \qquad $\triangleright$ Second step update  
			\State $x_{i}^{n+1} \gets x_{i}^{n}$ \label{code:l13} \qquad $\triangleright$Store parameter for next iteration
			\State \Call{UpdateTimeVector} {} \label{code:l14}
			\EndFor\\ \label{code:l15}
			\Return { $x_{i}$} \label{code:l16} 
			\Procedure{UpdateTimeVector}{} \label{code:l17}
			\State $t_{i,m} \gets$ Recorded iteration time \label{code:l18}
			\State $\boldsymbol{T}_{i}[m] \gets \beta \boldsymbol{T}_{i}[m]+(1-\beta)t_{i,m}$  \label{code:l19}
			\EndProcedure \label{code:l20}
		\end{small}
	\end{algorithmic}
\end{algorithm}

To obtain the communication policy that efficiently reduces the convergence time, we formulate and solve an optimization problem. Our objective is to minimize the total convergence time $k\overline{t}$, as shown in Eq. (\ref{eq:o0}).
\begin{figure}[h]
	\centering 
	\begin{equation} 
	\quad\min \quad k \overline{t} \label{eq:o0}
	\end{equation}
	\begin{flushleft}
		\qquad \qquad s.t.
		\begin{equation}
		\lambda^k \leq \varepsilon, \forall \varepsilon > 0 \label{eq:o1}
		\end{equation}
		\begin{equation}
		\overline{t} = \frac{\overline{t}_{i}}{M}= \frac{1}{M}\sum\nolimits_{m=1}^{M}t_{i,m}p_{i,m}d_{i,m}, \forall i\in [M] \label{eq:o2}
		\end{equation}
		\begin{equation}
		p_{i,m} > \alpha\rho(d_{i,m}+d_{m,i}),\forall i,m \in [M], m\neq i, d_{i,m}\neq 0\label{eq:o3}
		\end{equation}
		\begin{equation}
		p_{i,m} = 0, \forall i,m \in [M], d_{i,m}=0\label{eq:o4}
		\end{equation}
		\begin{equation}
		\sum\nolimits_{m=1}^{M}p_{i,m} = 1, \forall i \in [M] \label{eq:o5}
		\end{equation}
		
	\end{flushleft}
\end{figure}

\begin{algorithm}[tp] 
	\caption{Communication Policy Generation} 
	\label{alg:get-policy} 
	\begin{algorithmic}[1] %
		\begin{small}
			\Require learning rate $\alpha$, outer-loop search round $K$, inner-loop search round $R$, iteration time matrix $\boldsymbol{T}=[t_{i,m}]_{M\times M}$.
			\Ensure	Probability matrix $\boldsymbol{P}$
			\Function{GeneratePolicyMatrix}{$\alpha,K,R,\boldsymbol{T}$}\label{code:p1}
			\State  $L_{\rho} \gets 0.0$ \label{code:p2}
			\State  $U_{\rho} \gets 0.5/\alpha$ \label{code:p3}
			\State $\delta_{\rho} \gets (U_{\rho}-L_{\rho})/K$;\label{code:p4}
			\For {$k \in \{1,2,\dots,K\}$} \label{code:p5}
			\State $\rho \gets L_{\rho}+\delta_{\rho}$ \label{code:p6}
			\State $\boldsymbol{P},T_{convergence}\gets$ \Call{InnerLoop}{$\alpha,\rho,R,\boldsymbol{T}$} \label{code:p7}
			\State $out[k]\gets (\boldsymbol{P},T_{convergence},\rho)$\label{code:p8}
			\EndFor \label{code:p9}
			\State $y\gets$ find item in $out$ with minimum $T_{convergence}$\label{code:p10}
			\State \Return $y.\boldsymbol{P}$, $y.\rho$ \label{code:p11}
			\EndFunction \label{code:p12}
			\Function{Innerloop}{$\alpha,\rho,R,\boldsymbol{T}$} \label{code:p13}
			\State  $L \gets \max_{i\in [M]}\{\frac{\alpha\rho}{M}\sum_{i=m}^{M}t_{i,m}(d_{i,m}+d{m,i})\}$;
			\label{code:p14}
			\State  $U \gets \min_{i\in[M]}\frac{1}{M}\max_{m \in [M]}\{t_{i,m}d_{i,m}\}$; \label{code:p15}
			\State $\delta \gets (U-L)/R$;\label{code:p16}
			\For{$r \in \{1,2,\dots,R\}$} \label{code:p17}
			\State $\overline{t} \gets L+\delta$ \label{code:p18}  
			\State $\boldsymbol{P}\gets$ Solve LP by using standard method \label{code:p19}
			\State $\lambda_2 \gets$ Compute second largest eigenvalue for $\boldsymbol{Y_p}$ \label{code:p20}
			\State $T_{convergence} \gets \overline{t}\frac{\ln\varepsilon}{\ln\lambda_{2}}$ \label{code:p21}
			\State $res[r]\gets (T_{convergence},\lambda_{2},\overline{t},\boldsymbol{P})$
			\label{code:p22}
			\EndFor \label{code:p23}
			\State $x\gets$ find item in $res$ with minimum $T_{convergence}$\label{code:p24}	
			\State \Return $x.\boldsymbol{P}$, $x.T_{convergence}$ \label{code:p25}
			\EndFunction \label{code:p26}
		\end{small}
	\end{algorithmic}
\end{algorithm}

From Eq. (\ref{eq:j2}) we know that when $\lambda^k$ is less than a very small positive value $\varepsilon$, the models of worker nodes converge to a very small domain near the optimal model. Therefore, $k$ can be constrained by $\lambda$, as shown in Eq. (\ref{eq:o1}), where a small $\lambda$ means a small convergence iteration $k$ and, thus, a fast convergence rate. 
Constraint Eq. (\ref{eq:o2}) and (\ref{eq:o3}) ensure that the matrix $\mathbb{E}[(\boldsymbol{D}^k)^{T} \boldsymbol{D}^k]$ is a doubly stochastic symmetric matrix and its $\lambda$ is strictly less than 1. 
Consequently, constraint Eq. (\ref{eq:o1}) always holds. 
Constraint Eq. (\ref{eq:o2}) defines the global average iteration time $\overline{t}$, where $\overline{t}_{i}$ is defined in Eq. (\ref{eq:d3}). Eq. (\ref{eq:o2}) is derived by letting the rows of $\mathbb{E}[(\boldsymbol{D}^k)^{T} \boldsymbol{D}^k]$ sum to 1.
Constraint Eq. (\ref{eq:o3}) is derived by letting the entry $y_{i,m}$ of $\mathbb{E}[(\boldsymbol{D}^k)^{T} \boldsymbol{D}^k]$ to be larger than 0, for all $d_{i,m}\neq 0$, which determines the smallest communication probabilities between neighbors. 
These smallest communication probabilities are proportional to the adjustable coefficient $\rho$ given learning rate $\alpha$ and communication graph $\mathcal{\boldsymbol{G}}$. 
Constraints Eq. (\ref{eq:o4}) and (\ref{eq:o5}) naturally hold.

Eq. (\ref{eq:o0})-(\ref{eq:o5}) represent a nonlinear programming problem involving the computation of eigenvalues, which is hard to be solved by standard methods. Instead of finding its optimal solution, we propose a faster method to find a feasible sub-optimal solution. Our observation is that when given $\rho$ and $\overline{t}$, there is a feasible policy $\boldsymbol{P}$ that minimizes the objective value, $k\overline{t}$.  
Hence, our key idea is to search for a policy $\boldsymbol{P}$ that has the smallest objective value under various feasible configurations of $\rho$ and $\overline{t}$. 

\autoref{alg:get-policy} describes our method to find a feasible policy, which consists of two nested loops. The outer-loop searches for $K$ values of $\rho$ within its feasible interval $[L_{\rho},U_{\rho}]$ ({lines \ref{code:p1}-\ref{code:p6}}). For any given $\rho$, we use an inner-loop to obtain a sub-optimal matrix $\boldsymbol{P}$ ({line \ref{code:p7}}). At the end, we return the matrix $\boldsymbol{P}$ with the smallest objective value ({lines \ref{code:p10}-\ref{code:p11}}). The inner-loop searches for $R$ values of $\overline{t}$ within its feasible interval $[L,U]$ when $\rho$ is given ({lines~\ref{code:p14}-\ref{code:p18}}). For each value of $\overline{t}$ in the inner-loop, we solve a linear programming (LP) problem defined in Eq. (\ref{eq:lp0}) to obtain a sub-optimal matrix $\boldsymbol{P}$ ({lines \ref{code:p19}-\ref{code:p22}}). \par
	\begin{equation} 
	(LP)~\min\sum\nolimits_{i=1}^{M}p_{i,i}~s.t.~Eq.(\ref{eq:o2}), Eq.(\ref{eq:o3}), Eq.(\ref{eq:o4}), Eq.(\ref{eq:o5})
	\label{eq:lp0}
	\end{equation}

The LP problem in Eq. (\ref{eq:lp0}) seeks to minimize the probabilities of distributed workers to select themselves. Such a design encourages the workers to exchange information with their neighbors and, thus, achieve a high convergence rate.
\ifthenelse{\boolean{isfullpaper}}
{
\color{black}
 The derivation of the feasible intervals for $\rho$ and $\overline{t}$ can be found in Appendix~\ref{apendex:a}. \autoref{alg:get-policy} can solve the problem stated in Eq. (\ref{eq:o0})-(\ref{eq:o5}) with an approximation ratio of $\frac{U}{L}\frac{\ln(M\!-\!1)\!-\!\ln(M\!-\!3)}{\ln(1\!-\!2a\!+\!a^{M})\!-\!\ln(1\!-\!2a\!+\!a^{M\!+\!1})}$ when the graph connecting the worker nodes is fully-connected, the network is heterogeneous, and where $a$ denotes the minimum positive entry of $\mathbb{E}[(\boldsymbol{D}^k)^{T} \boldsymbol{D}^k]$. The proof can be found in Appendix~\ref{apendex:a-1}.
}
{
 The derivation of the feasible intervals for $\rho$ and $\overline{t}$ can be found in Appendix A of the \fullpaperterm{}.
 \color{black}
 \autoref{alg:get-policy} can solve the problem stated in Eq. (\ref{eq:o0})-(\ref{eq:o5}) with an approximation ratio of $\frac{U}{L}\frac{\ln(M\!-\!1)\!-\!\ln(M\!-\!3)}{\ln(1\!-\!2a\!+\!a^{M})\!-\!\ln(1\!-\!2a\!+\!a^{M\!+\!1})}$ when the graph connecting the worker nodes is fully-connected, the network is heterogeneous, and where $a$ denotes the minimum positive entry of $\mathbb{E}[(\boldsymbol{D}^k)^{T} \boldsymbol{D}^k]$. The proof can be found in the \fullpaperterm{}.
}



\ifthenelse{\boolean{isfullpaper}}
{
\color{black}
}
{
\color{black}
}
\subsection{Extension to Existing Decentralized PSGD Approaches}
\label{subsec:netmax-extension}
A noteworthy aspect is that the standalone \coordinator{} of \ourtech{} can be used to improve existing D-PSGD approaches (e.g., Gossiping SGD~\cite{jin2016scale} and AD-PSGD~\cite{lian2018asynchronous}) as long as they can be re-written as the form of Eq.~(\ref{eq:j1}).
Specifically, the extension includes two steps.
First, we can derive a similar constraint to replace Eq.~(\ref{eq:o3}) by requiring each entry $y_{i,m}$ of $\mathbb{E}[(\boldsymbol{D}^k)^{T} \boldsymbol{D}^k]$ to be larger than $0$, for all $d_{i,m}\neq 0$. 
Second, the corresponding probabilities for worker nodes to choose neighbors can be obtained by solving the optimization problem Eq.~(\ref{eq:o0})-(\ref{eq:o5}).
By doing so, existing D-PSGD approaches can be enhanced to adapt to the dynamics of heterogeneous networks. 
\ifthenelse{\boolean{isfullpaper}}
{
}
{}

\color{black}
\section{Convergence analysis}
\label{sec:convergence-analysis}

In this section, we present the convergence analysis of our proposed approach. More specifically, we prove that the models of distributed worker nodes converge to the same optima under Assumption 1.

\noindent
\textbf{Assumption 1.} We make the following commonly used assumptions for this analysis:
\begin{itemize}
	\item \textbf{Connected graph}. The undirected graph $\mathcal{\boldsymbol{G}}$ connecting worker nodes is connected.
	\item \textbf{Lipschitzian gradient}. The local loss function $f$ is a $\mu$-strongly convex with $L$-$Lipschitz$ gradients.
	\item {
	\ifthenelse{\boolean{isfullpaper}}
    {}
    {
     \color{black}
    }
	\textbf{Bounded gradient variance.} For any worker node $i$, its gradient can be bounded as $\mathbb{E} [\nabla f(x_{i}^{k})^{T}\nabla f(x_{i}^{k})] \leq \eta^2$.}
	\item \textbf{Additive noise}. For any worker node $i$, its stochastic gradients have additive noise $\xi_i$ caused by stochastic sampling from the dataset. $\xi_i$ is with zero mean $\mathbb{E}[\xi_i]=0$ and bounded variance $\mathbb{E}[\xi_{i}^{T} \xi_{i}] \leq \sigma^{2}$.
\end{itemize}

For simplicity, let $\gamma_{i,m} =\frac{d_{i,m}+d_{m,i}}{2p_{i,m}} $ and $g_{i}^{k} = \nabla f(x_{i};\mathcal{D}_i)+\xi_i^k$, where $\xi_i^k$ is the additive noise at global step $k$.

At global step $k$, worker node $i$ pulls parameters from its neighbor $m$ with probability $p_{i,m}$. According to \autoref{alg:nadsgd}, the first update of worker node $i$ using local gradients ({line \ref{code:l9}}) can be written as
\begin{align}
x_{i}^{k+\frac{1}{2}}& = x_{i}^{k} - \alpha g_{i}^{k} \label{eq:c0}
\end{align}
The second update of worker node $i$ using neighbor's parameters ({lines \ref{code:l11}-\ref{code:l13}}) can be written as
\begin{flalign}
x_{i}^{k+1} &=x_{i}^{k+\frac{1}{2}} -  \alpha \rho\gamma_{i,m}\Big(x_{i}^{k+\frac{1}{2}} - x_{m}^k\Big)\nonumber\\
&= (1-\alpha\rho\gamma_{i,m})\Big(x_{i}^{k} - \alpha g_{i}^{k}\Big)+\alpha\rho\gamma_{i,m}  x_{m}^k \label{eq:c1}
\end{flalign}
For any worker node $j \neq i$, $x_{j}^{k+1} = x_{j}^{k}$. Thus, at each global step $k$, we have
\begin{equation}
\left\{
\begin{array}{lr}
x_{i}^{k+1}=(1-\alpha\rho\gamma_{i,m})\Big(x_{i}^{k} - \alpha  g_{i}^{k}\Big)+\alpha\rho\gamma_{i,m}  x_{m}^k  \\
x_{j}^{k+1} = x_{j}^{k},  \ \forall j\in[M], \ j \neq i 
\end{array}
\right. \label{eq:c2}
\end{equation}
We can rewrite Eq. (\ref{eq:c2}) in matrix form as \begin{equation} 
\boldsymbol{x}^{k+1}=\boldsymbol{D}^{k}(\boldsymbol{x}^{k}-\alpha \boldsymbol{g}^{k})  \label{eq:c3}
\end{equation}
where
$\boldsymbol{x}^{k}=\left[x_{1}^{k},\dots, x_{i}^{k},\dots, x_{m}^{k},\dots, x_{M}^{k} 
\right]^T$ is a vector containing all worker nodes’ models at global step $k$, $\boldsymbol{g}^{k}=\left[
0,\dots,g_{i}^{k},\dots,0,\dots,0 
\right]^T$ is a vector containing all worker nodes’ gradients at global step $k$ and $\boldsymbol{D}^{k}$ is an $M\times M$ matrix, which is expressed as\\
\begin{equation}
\boldsymbol{D}^{k}= \boldsymbol{I}+\alpha\rho\gamma_{i,m} \boldsymbol{e}_i (\boldsymbol{e}_m - \boldsymbol{e}_i)^T \label{eq:c4}
\end{equation}
where $\boldsymbol{e}_i=\left[0,\dots,1,\dots,0,\dots,0\right]^T$ is an $M\times1$ unit vector with $i$-th component equal to 1.
Note that index $i$ is a random variable drawn from $\{1,2,\dots,M\}$ with probability $p_i$ and $m$ is a random variable drawn from $\{1,2,\dots,M\}$ with probability $p_{i,m}$. The expectation below is derived with respect to the random variables $i$ and $m$:
\begin{flalign}
\boldsymbol{Y} &= \mathbb{E}\Big[(\boldsymbol{D}^k)^{T} \boldsymbol{D}^k\Big] &\nonumber \\
&= \mathbb{E}\Big[\boldsymbol{I} \!+\! \alpha\rho\gamma_{i,m} \boldsymbol{e}_{i}(\boldsymbol{e}_{m}-\boldsymbol{e}_{i})^{T}\!+\alpha\rho\gamma_{i,m}\big[\boldsymbol{e}_{i}(\boldsymbol{e}_{m}-\boldsymbol{e}_{i})^{T}\big]^{T} &\nonumber\\
&\qquad+\alpha^{2}\rho^{2}\gamma_{i,m}^{2}\big[\boldsymbol{e}_{i}(\boldsymbol{e}_{m}-\boldsymbol{e}_{i})^{T}\big]^{T} \boldsymbol{e}_{i}(\boldsymbol{e}_{m}-\boldsymbol{e}_{i})^{T} \Big]& \label{eq:d0}\\
&=\left[y_{i,m}\right]_{M\times M} &\label{eq:d1}
\end{flalign}
where
\begin{equation}
\left\{
\begin{array}{lr}
y_{i,i}=1-2\alpha\rho\sum_{\substack{m\in [M]\\ m\neq i}}p_{i}p_{i,m}\gamma_{i,m}&\\
\ 
\qquad +\alpha^{2}\rho^{2}\sum_{\substack{m\in [M]\\ m\neq i}}(p_{i}p_{i,m}\gamma_{i,m}^{2}\!+\!p_{m}p_{m,i}\gamma_{m,i}^{2}), \forall i\in [M]& \\
&\\  
y_{i,m} = \alpha\rho(p_{i}p_{i,m}\gamma_{i,m}+p_{m}p_{m,i}\gamma_{m,i})&\\
\ \qquad-\alpha^{2}\rho^{2}(p_{i}p_{i,m}\gamma_{i,m}^{2}\!+\!p_{m}p_{m,i}\gamma_{m,i}^{2}),\forall i,m\!\in\! [M],m\!\neq\! i & 
\end{array}
\right. \label{eq:d2}
\end{equation}

The matrix $\boldsymbol{Y}$ is associated with the communication policy matrix $\boldsymbol{P}$, denoted by $\boldsymbol{Y_P}$. Let $\lambda_{1}$ and $\lambda_{2}$ be the largest and second largest eigenvalue of matrix $\boldsymbol{Y_P}$, respectively. Let $\lambda=\lambda_{2}$ if $\boldsymbol{Y_P}$ is a doubly stochastic matrix; otherwise let $\lambda=\lambda_{1}$. Then we can derive the following two theorems under static and dynamic networks. 
\ifthenelse{\boolean{isfullpaper}}
{
The proofs can be found in Appendix~\ref{apendex:b} and Appendix~\ref{apendex:c}, respectively.
}
{
The proofs can be found in Appendix C and Appendix D of the \fullpaperterm{}.
}

\begin{theorem}
\label{theorem:static}
(static network) If Assumption 1 is satisfied and the applied decentralized learning algorithm can be written in the form of Eq. (\ref{eq:c3}) with learning rate $0<\alpha\leq \frac{2}{\mu + L}$, the squared sum of the local model parameters’ deviation from the optimal $x^{*}$ can be bounded as\\
	\begin{equation}
	\mathbb{E}[ \lVert \boldsymbol{x}^{k}-x^{*}\boldsymbol{1}\rVert^{2}] \leq \lambda^{k} \lVert \boldsymbol{x}^{0}-x^{*}\boldsymbol{1} \rVert^{2}+ \alpha^{2}\sigma^{2} \frac{\lambda}{1-\lambda} \label{eq:t1}
	\end{equation}
	The decentralized training will converge to a small domain and achieve consensus if $\lambda<1$.
\end{theorem}

In a dynamic network, when network conditions change, so does $\lambda$. Let $\lambda_{max}$ be the maximum among historical eigenvalues $\lambda$. Then we can formulate Theorem~\ref{theorem:dynamic}.

\begin{theorem}
\label{theorem:dynamic}
(dynamic network) If Assumption 1 is satisfied and the applied decentralized learning algorithm can be written in the form of Eq. (\ref{eq:c3}) with a learning rate $0<\alpha\leq \frac{2}{\mu + L}$, the squared sum of the local model parameters’ deviation from the optimal $x^{*}$ can be bounded as
\\
	\begin{equation}
	\mathbb{E}[ \lVert \boldsymbol{x}^{k}-x^{*}\boldsymbol{1}\rVert^{2}] \leq \lambda_{max}^{k} \lVert \boldsymbol{x}^{0}-x^{*}\boldsymbol{1} \rVert^{2}+ \alpha^{2}\sigma^{2} \frac{\lambda_{max}}{1-\lambda_{max}}\label{eq:t2}
	\end{equation}
	The decentralized training will converge to a small domain and achieve consensus if $\lambda_{max}<1$.
\end{theorem}

From Theorem~\ref{theorem:static} and Theorem~\ref{theorem:dynamic}, we obtain that the communication policy can determine the eigenvalues of $\boldsymbol{Y_P}$ under any given network conditions, and, thus, determine the convergence of the training. The convergence rate can be bounded by $\lambda^k$ and $\lambda_{max}^{k}$ under static and dynamic networks, respectively. Moreover, we can formulate the following theorem.

\begin{theorem}
\label{theorem:converge}
For any feasible solution $\boldsymbol{P}$ that is obtained by \autoref{alg:get-policy}, $\boldsymbol{Y_P}$ is a doubly stochastic matrix and the second largest eigenvalue $\lambda$ of matrix $\boldsymbol{Y_P}$ is strictly less than 1. Hence, the decentralized training converges. 
{
Moreover, for a chosen learning rate $\alpha=\frac{c}{\sqrt{k}}$ where $c$ is a pre-defined positive number, the distributed learning converges with a rate $O(1/\sqrt{k})$.}
\end{theorem}

Any feasible solution of the optimization problem in Eq. (\ref{eq:o0})-(\ref{eq:o5}) ensures that the training converges. As the policy $\boldsymbol{P}$ obtained by \autoref{alg:get-policy} is always a feasible solution for the optimization problem in Eq. (\ref{eq:o0})-(\ref{eq:o5}), the training performed by our approach always converges. 
\ifthenelse{\boolean{isfullpaper}}
{
The detailed proof of Theorem~\ref{theorem:converge} is presented in Appendix~\ref{apendex:d}.
}
{
The proof of Theorem~\ref{theorem:converge} is presented in Appendix E of the \fullpaperterm{}.
}

\section{Performance Evaluation}
\label{sec:eval}

We have implemented \ourtech{} in Python on top of PyTorch. 
In this section, we conduct experiments to evaluate \ourtech{} against three state-of-the-art decentralized machine learning (ML) techniques: \allreduce{}~\cite{jia2018highly}, \adpsgd{}~\cite{lian2018asynchronous} and \prague{}~\cite{luo2020prague}.
Note that, for a fair comparison, all the above implementations share the same runtime environment. 

\subsection{Experimental Setup}
\label{subsec:exp-setup}

\noindent\textbf{Cluster.} 
Experiments are conducted in a multi-tenant cluster consisting of $18$ servers connected via $1000$ Mbps Ethernet. 
Each server is equipped with an Intel(R) Xeon(R) W-2133 CPU @ 3.60GHz, $65$ GB DDR3 RAM, and $3$ or $8$ GPUs of GeForce RTX 2080 Ti.  
Worker nodes and \coordinator{} of \ourtech{} are run in docker containers, each of worker nodes is assigned with an exclusive GPU. 
Due to the limited number of GPUs available on each server, we run $4$, $8$ and $16$ worker nodes across $2$, $3$ and $4$ servers, respectively. 
Unless otherwise specified, the number of worker nodes is $8$ by default throughout the experiments. 

\noindent\textbf{Network.}
Some links among worker nodes are purposely slowed down in order to emulate a dynamic heterogeneous network.
In particular, by following \cite{lian2018asynchronous}, we randomly slow down one of the communication links among nodes by 2x to 100x. 
Other than that, we further change the slow link every $5$ minutes. 
As the processing time of distributed ML jobs varies from tens of minutes to several days~\cite{lian2018asynchronous, jeon2019analysis}, the above frequency of link speed change is appropriate to mimic the dynamic arriving/leaving of distributed ML jobs in a multi-tenant cluster. 

On the other hand, to emulate a homogeneous network, we reserve a server with $8$ GPUs and scale the number of worker nodes running within the server from $4$ to $8$. 
The worker nodes are connected via a virtual switch~\cite{ovs} with $10$ Gbps bandwidth, which implements a homogeneous peer-to-peer communication.

\ifthenelse{\boolean{isfullpaper}}
{}
{
\color{black}
}
\noindent\textbf{Datasets and models.} 
We evaluate the experiments using five real-world datasets: MNIST, \cifar{}, CIFAR100, Tiny-ImageNet, and ImageNet. We use four models in the experiments: MobileNet, \resnet{}, ResNet50, and \vgg{} whose numbers of parameters are approximately 4.2M, 11.7M, 25.6M, and 143.7M, respectively.
\color{black}

\noindent\textbf{Configurations.}
The hyper-parameter settings as applied in both \adpsgd{}~\cite{lian2018asynchronous} and \prague{}~\cite{luo2020prague} are adopted in our experimental study. 
Specifically, unless noted otherwise, the models are trained with \emph{batch size} $128$, \emph{momentum} $0.9$, and \emph{weight decay} $10^{-4}$. 
The \emph{learning rate} starts from $0.1$ and decays by a factor of $10$ once the loss does not decrease any more. 
We train \resnet{} and \vgg{} with \emph{epoch numbers} of $64$ and $82$ respectively. 
For \ourtech{}, we alter the communication policy every $T_s = 2$ minutes, as such a period is long enough for the detection of link speed change at runtime. 

\subsection{Performance of Decentralized Training}
\label{subsec:exp-performance}

\begin{figure}[tp]
	\centering 
	\begin{minipage}[b]{0.23\textwidth}
		\centering
		\subfigure[\resnet{}] { \label{fig:epoch-time:a}     \includegraphics[width=1\textwidth]{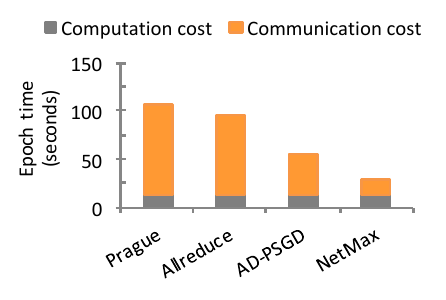} }
		\vspace{-2ex}
	\end{minipage}
	\hspace{0.01in}
	\begin{minipage}[b]{0.23\textwidth}
		\centering 
		\subfigure[\vgg{}] { \label{fig:epoch-time:b}     \includegraphics[width=1\textwidth]{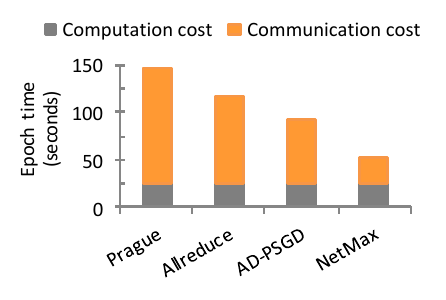} }  
		\vspace{-2ex} 
	\end{minipage} 
	\caption{Average epoch time with 8 worker nodes in a heterogeneous network.}
	\label{fig:epoch-time}
	\vspace{-1ex}
\end{figure}

\begin{figure}[tp]
	\centering 
	\begin{minipage}[b]{0.23\textwidth}
		\centering
		\subfigure[\resnet{}] { \label{fig:epoch-time-homo:a}     \includegraphics[width=1\textwidth]{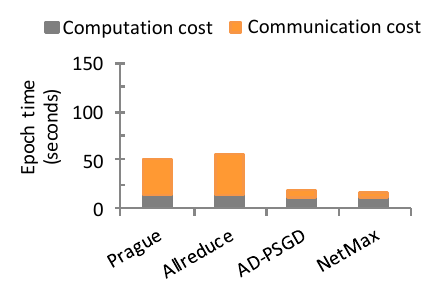} }
		\vspace{-2ex}
	\end{minipage}
	\hspace{0.01in}
	\begin{minipage}[b]{0.23\textwidth}
		\centering 
		\subfigure[\vgg{}] { \label{fig:epoch-time-homo:b}     \includegraphics[width=1\textwidth]{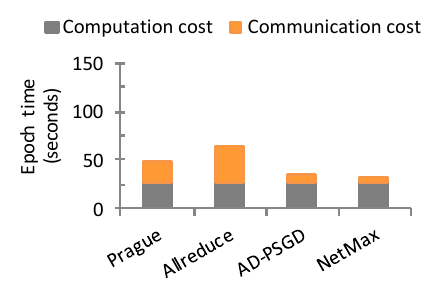} } 
		\vspace{-2ex} 
	\end{minipage} 
	\caption{Average epoch time with 8 worker nodes in a homogeneous network.}
	\label{fig:epoch-time-homo}
	\vspace{-1ex}
\end{figure}

\autoref{fig:epoch-time} presents the computation cost and communication cost of the average epoch time of the four approaches under the heterogeneous network setting. 
As can be seen, the computation costs of all the approaches are almost the same. 
This is expected since they train the same models under the same runtime environment. 
However, the communication costs differ significantly. 
In particular, \ourtech{} incurs the lowest communication cost, mainly due to its improved utilization of high-speed links. 
For \resnet{}, \ourtech{} reduces the communication cost by up to $83.4\%$, $81.7\%$ and $63.7\%$ comparing with \prague{}, \allreduce{}, and \adpsgd{}, respectively. 
For \vgg{}, \ourtech{} reduces the communication cost by up to $76.8\%$, $69.2\%$ and $58.6\%$, correspondingly.  
In contrast, \prague{} suffers from the highest communication cost. 
This is because \prague{} divides the worker nodes into multiple groups in every iteration, and each group executes a \textit{partial-allreduce} operation to average the models.
The concurrent executions of partial-allreduce of different groups compete for the limited bandwidth capacity, resulting in network congestion. 
Moreover, the partial-allreduce operation is agnostic to the link speed. 
As a consequence, the heavy use of low-speed links further deteriorates its performance.

We repeat the above experiments but alternatively under the homogeneous network setting, and the results are shown in \autoref{fig:epoch-time-homo}. 
As can be seen, the computation costs of the four approaches remain similar, whereas the communication costs are fairly lower comparing with \autoref{fig:epoch-time}. 
This is due to the expanded bandwidth used in the homogeneous network. 
Overall, \ourtech{} and \adpsgd{} are superior in low communication cost over \prague{} and \allreduce{}. 
This is because \allreduce{} and \prague{} average gradients or model parameters from multiple worker nodes in each iteration, which introduces extra rounds of communication between worker nodes and thus results in high communication cost. 
In contrast, each worker node in \ourtech{} and \adpsgd{} pulls the model updates from only one neighbor and immediately starts the next iteration once that communication completes, leading to the reduction of communication cost.

\begin{figure}[!t]
	\centering
	\includegraphics[width=0.46\textwidth]{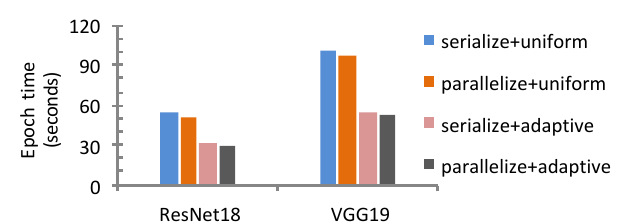}
	\caption{Average epoch time \textit{w.r.t.} the four settings.}
	\label{fig:serialize-parallel} 
	\vspace{-1ex}
\end{figure}

{
\ifthenelse{\boolean{isfullpaper}}
{}
{
\color{black}
}
\subsection{Source of Performance Improvement in \ourtech{}}
\label{subsec:probability-performance}
Two strategies are implemented in \ourtech{} to reduce the epoch time, as elaborated in \autoref{alg:nadsgd}. 
First, the gradient computation and network communication are executed in parallel. 
Second, each worker node selects its neighbors based on the adaptive probabilities. 
In this subsection, we evaluate the performance improvement brought about by these two strategies when training the ResNet18 and VGG19 models on the CIFAR10 dataset.
In particular, we analyze the following four settings.
\begin{itemize}[leftmargin=*]
    \item Setting 1: serial execution (of gradient computation and network communication) + uniform probabilities (for selecting neighbors). This represents the baseline.
    \item Setting 2: parallel execution + uniform probabilities. This is to verify the effect of parallelism.
    \item Setting 3: serial execution + adaptive probabilities. This is to verify the effect of adaptive probabilities.
    \item Setting 4: parallel execution + adaptive probabilities. This represents the full functioning \ourtech{}.
\end{itemize}

As shown in \autoref{fig:serialize-parallel}, the use of adaptive probabilities contributes to the majority of performance gain in \ourtech{}. For example, when comparing serial+uniform with serial+adaptive, we observe that the average epoch times are reduced from 54s and 100.5s to 30.3s and 55.4s for ResNet18 and VGG19, respectively. On the other hand, the performance improvement due to parallelization is marginal because the time of gradient computation on GPUs is much shorter than the time of network communication.
}

\subsection{Convergence and Accuracy}
\label{subsec:exp-convergence}

\begin{figure}[tp]
	\centering 
	\begin{minipage}[b]{0.23\textwidth}
		\centering
		\subfigure[\resnet{}] { \label{fig:loss-time:a}     \includegraphics[width=1\textwidth]{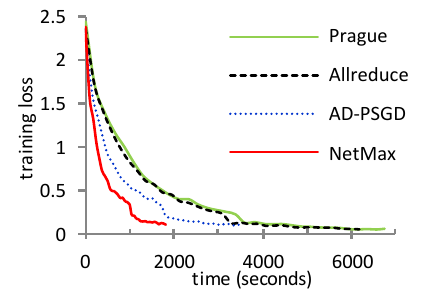} }
		\vspace{-2ex}
	\end{minipage}
	\hspace{0.01in}
	\begin{minipage}[b]{0.23\textwidth}
		\centering 
		\subfigure[\vgg{}] { \label{fig:loss-time:b}     \includegraphics[width=1\textwidth]{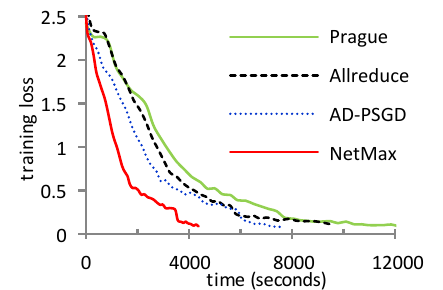} }  
		\vspace{-2ex} 
	\end{minipage} 
	\caption{Training loss with 8 worker nodes in a heterogeneous network.
	}
	\label{fig:loss-time}
	\vspace{-1ex}
\end{figure}

\begin{table}[tp] \scriptsize
	\newcommand{\multirow}
	\centering
	\caption{Accuracy of models trained over a heterogeneous network}
	\begin{tabular}{ll|l|l|l|l}
		\hline
		&&\prague{}&Allreduce&\adpsgd{}&\ourtech{}\\
		\hline
		                         &4 nodes&89.56\%&90.11\%&90.24\%&90.46\%\\	
	                     \resnet{}&8 nodes&90.54\%&90.16\%&90.39\%&91.14\%\\
		                         &16 nodes&90.44\%&90.37\%&90.23\%&90.78\%\\
		\hline
		                         &4 nodes&90.09\%&89.98\%&89.8\%&90.24\%\\	
		                 \vgg{}&8 nodes&89.76\%&89.05\%&90.47\%&90.74\%\\
		                         &16 nodes&90.14\%&90.13\%&90.22\%&91.07\%\\
		\hline
	\end{tabular}
	\label{tab:acc-hetero}
	\vspace{-1ex}
\end{table}

\autoref{fig:loss-time} shows the comparison of training loss with respect to training time under the heterogeneous network setting. 
As can be seen, \ourtech{} converges much faster than the other approaches. 
For \resnet{}, \ourtech{} achieves about 3.7$\times$, 3.4$\times$ and 1.9$\times$ speedup over \prague{}, \allreduce{} and \adpsgd{}, respectively. 
For \vgg{}, the speedup is about 2.8$\times$, 2.2$\times$ and 1.7$\times$, correspondingly. 
Such performance gain is rooted at \ourtech{}'s capability such that it can adapt the probabilities for neighbor selection with respect to the change of link speed to fine-tune the communication frequencies between worker node pairs (referring to the policy design in \autoref{subsection:plicy-design}).
This not only allows the worker nodes to communicate preferably over high-speed links, but also achieves a fast convergence rate, bringing about the decrement of the overall convergence time in \ourtech{}.

\autoref{tab:acc-hetero} summarizes the test accuracy of all the approaches corresponding to the aforementioned convergences. 
The result shows that all the approaches can achieve around $90\%$ test accuracy for both \resnet{} and \vgg{}, while \ourtech{} performs slightly better than the others. 
Such an accuracy gain of \ourtech{} is attributed to the adjustable probabilities for worker nodes to select neighbors introducing more randomness in the distributed training, which could benefit the training to escape from a shallow and sharp sub-optima that leads to inferior generalization~\cite{neelakantan2017adding,kleinberg2018an,keskar2017on}.

\begin{figure}[tp]
	\centering 
	\begin{minipage}[b]{0.23\textwidth}
		\centering
		\subfigure[\resnet{}] { \label{fig:loss-time-homo:a}     \includegraphics[width=1\textwidth]{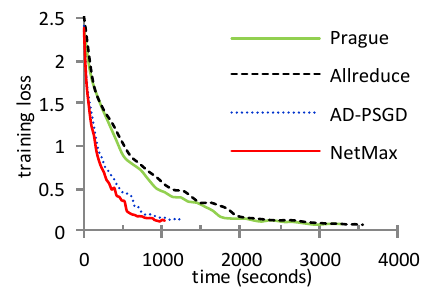} }
		\vspace{-2ex}
	\end{minipage}
	\hspace{0.01in}
	\begin{minipage}[b]{0.23\textwidth}
		\centering 
		\subfigure[\vgg{}] { \label{fig:loss-time-homo:b}     \includegraphics[width=1\textwidth]{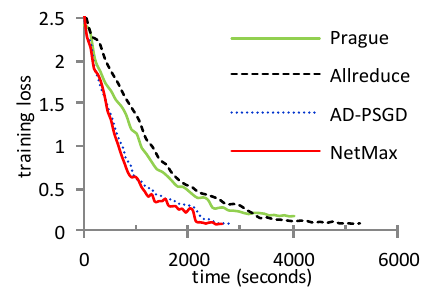} }  
		\vspace{-2ex} 
	\end{minipage} 
	\caption{Training loss with 8 worker nodes in a homogeneous network.}
	\label{fig:loss-time-homo}
	\vspace{-1ex}
\end{figure}

\begin{table}[tp]\scriptsize
	\newcommand{\multirow}
	\centering
	\caption{Accuracy of models trained over a homogeneous network}
	\begin{tabular}{ll|l|l|l|l}
		\hline
		&&\prague{}&Allreduce&\adpsgd{}&\ourtech{}\\
		\hline
		&4 nodes&89.37\%&89.45\%&89.83\%&90.19\%\\	
		\resnet{}&6 nodes&89.65\%&90.25\%&90.17\%&89.98\%\\
		&8 nodes&90.32\%&90.16\%&89.95\%&90.24\%\\
		\hline
		&4 nodes&89.97\%&90.41\%&90.69\%&90.62\%\\	
		\vgg{}&6 nodes&91.02\%&90.52\%&90.88\%&91.07\%\\
		&8 nodes&89.62\%&89.86\%&90.30\%&90.47\%\\
		\hline
	\end{tabular}
	\label{tab:acc-homo}
	\vspace{-1ex}
\end{table}

For the homogeneous network setting, \autoref{fig:loss-time-homo} exhibits that \ourtech{} is also advanced in the fastest convergence time. 
Note that \ourtech{} and \adpsgd{} have similar convergence trends. 
This is because the communication links in the homogeneous network share roughly equivalent link speed.
As a consequence, \ourtech{} lets worker nodes choose their neighbors randomly and uniformly to favor fast convergence, resulting in being similar to \adpsgd{} in terms of communication. 
In contrast, \allreduce{} and \prague{} converge much slower, since they both suffer from higher communication cost than \ourtech{} and \adpsgd{} as illustrated in \autoref{fig:epoch-time-homo}. 

Correspondingly, \autoref{tab:acc-homo} summarizes the test accuracy of all the approaches with respect to the above convergences, and the results are basically consistent with \autoref{tab:acc-hetero}. 

\subsection{Scalability}

\begin{figure}[tp]
	\centering 
	\begin{minipage}[b]{0.23\textwidth}
		\centering
		\subfigure[\resnet{}] { \label{fig:speedup:a}     \includegraphics[width=1\textwidth]{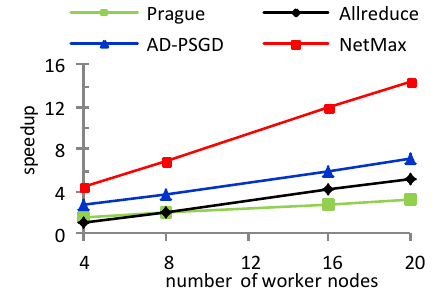} }
		\vspace{-2ex}
	\end{minipage}
	\hspace{0.01in}
	\begin{minipage}[b]{0.23\textwidth}
		\centering 
		\subfigure[\vgg{}] { \label{fig:speedup:b}     \includegraphics[width=1\textwidth]{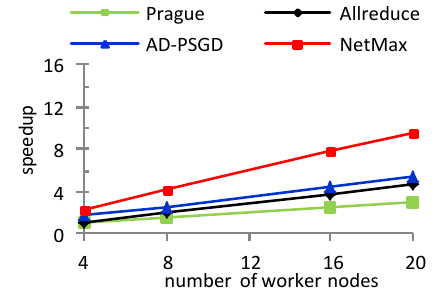} } 
		\vspace{-2ex} 
	\end{minipage} 
	\caption{Speedup \textit{w.r.t.} number of worker nodes in a heterogeneous network.}
	\label{fig:speedup}
	\vspace{-1ex}
\end{figure}

\begin{figure}[tp]
	\centering 
	\begin{minipage}[b]{0.23\textwidth}
		\centering
		\subfigure[\resnet{}] { \label{fig:speedup-homo:a}     \includegraphics[width=1\textwidth]{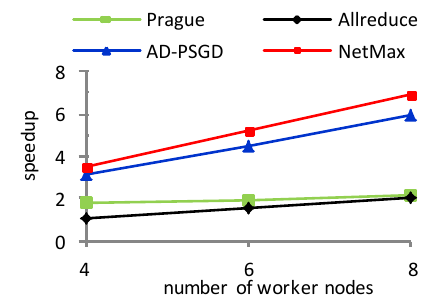} }
		\vspace{-2ex}
	\end{minipage}
	\hspace{0.01in}
	\begin{minipage}[b]{0.23\textwidth}
		\centering 
		\subfigure[\vgg{}] { \label{fig:speedup-homo:b}     \includegraphics[width=1\textwidth]{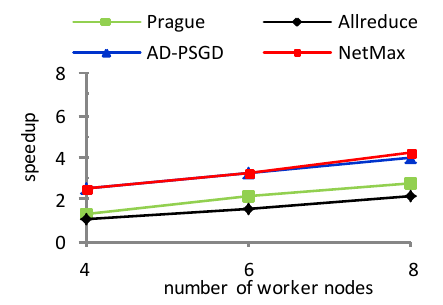} }  
		\vspace{-2ex} 
	\end{minipage} 
	\caption{
	Speedup \textit{w.r.t.} number of worker nodes in a homogeneous network.}
	\label{fig:speedup-homo}
	\vspace{-1ex}
\end{figure}

We evaluate the scalability of the four decentralized ML approaches. 
For ease of illustration, we set the training time after finishing a specified epoch (i.e., $64$ for \resnet{} and $82$ for \vgg{}) in \allreduce{} with $4$ worker nodes as the baseline, and refer to it to calculate the speedup of other runs. 
The results shown in \autoref{fig:speedup} and \autoref{fig:speedup-homo} indicate that \ourtech{} is superior in scalability over the other approaches under both the heterogeneous and homogeneous network settings. 
Moreover, the saving of communication cost in \ourtech{} further enlarges the performance gap along with the increment of nodes. 

{
\ifthenelse{\boolean{isfullpaper}}
{}
{
\color{black}
}
\subsection{Non-uniform Data Partitioning}
\label{subsec:nonuniform-data}
While the previous experiments are based on uniform data partitioning, in the following subsections, we evaluate \ourtech{} with non-uniform data partitioning. We train different models on various datasets with the following three settings.

\begin{itemize}[leftmargin=*]
    \item We train ResNet18 on CIFAR10, CIFAR100, and Tiny-ImageNet datasets, respectively, with 8 worker nodes instantiated in two GPU servers. Each server hosts 4 worker nodes, and each node occupies one GPU. Each dataset is partitioned into $10$ segments. The worker nodes on the first server, indexed by $\langle w0,w1,w2,w3 \rangle$, process one segment of data, respectively. The worker nodes on the second server, indexed by $\langle w4, w5, w6, w7 \rangle$, process $\langle 2,1,2,1 \rangle$ segments of data, respectively. 
    The batch size of each worker node is set to $64$ $\times$ the segment number. For example, the batch size of $w4$ is $64\times 2 = 128$. The learning rate starts from $0.1$ and decays by a factor of $10$ at epoch $80$. The total epoch number is set to $120$. 
    \item We train ResNet50 on ImageNet dataset with 16 worker nodes instantiated in two GPU servers. Each server hosts 8 worker nodes, and each node occupies one GPU. The dataset is partitioned into 20 segments. The worker nodes on the first server
    process one segment of data, respectively. The worker nodes on the second server
    process $\langle 2,1,2,1,2,1,2,1 \rangle$ segments of data, respectively. The batch size of each worker node is set to $64$ $\times$ the segment number. 
    The learning rate starts from $0.1$ and decays by a factor of $10$ at epoch $40$. The total epoch number is $75$.
    \item We train MobileNet on MNIST dataset with 8 worker nodes instantiated in two GPU servers. We consider an extreme condition where the worker nodes' data distributions are non-IID, as shown in \autoref{tab:lost-labels}. The batch size is set to $32$, and the learning rate is set to $0.01$.
\end{itemize}

\begin{figure}[!tp]
	\centering 
	\begin{minipage}[b]{0.23\textwidth}
		\centering
		\subfigure[Loss varies with epochs] { \label{fig:loss-cifar100:a}     \includegraphics[width=1\textwidth]{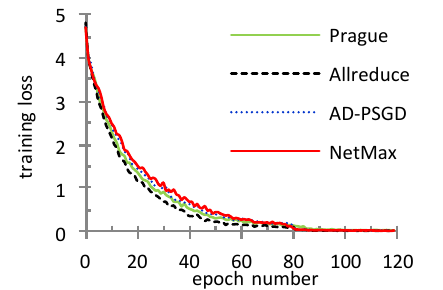} }
		\vspace{-2ex}
	\end{minipage}
	\hspace{0.01in}
	\begin{minipage}[b]{0.23\textwidth}
		\centering 
		\subfigure[Loss varies with time] { \label{fig:loss-cifar100:b}     \includegraphics[width=1\textwidth]{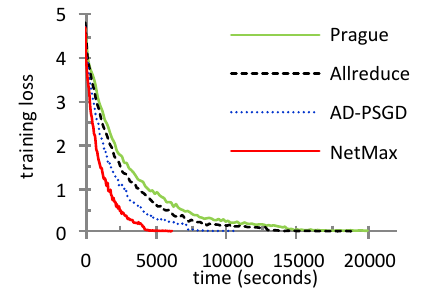} }  
		\vspace{-2ex} 
	\end{minipage} 
	\caption{Training ResNet18 on CIFAR100.}
	\label{fig:loss-cifar100}
	\vspace{-1ex}
\end{figure}

\begin{figure}[!tp]
	\centering 
	\begin{minipage}[b]{0.23\textwidth}
		\centering
		\subfigure[Loss varies with epochs] { \label{fig:loss-imagenet:a}     \includegraphics[width=1\textwidth]{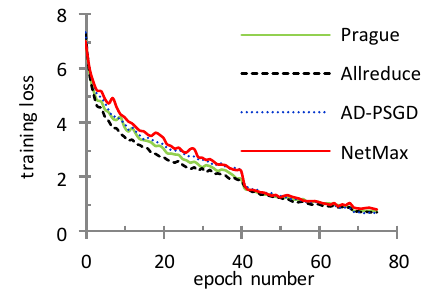} }
		\vspace{-2ex}
	\end{minipage}
	\hspace{0.01in}
	\begin{minipage}[b]{0.23\textwidth}
		\centering 
		\subfigure[Loss varies with time] { \label{fig:loss-imagenet:b}     \includegraphics[width=1\textwidth]{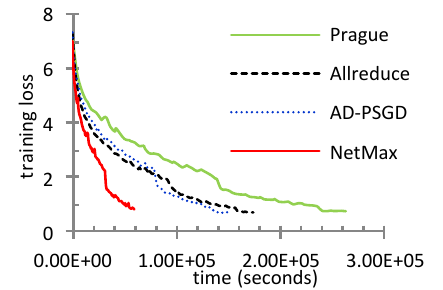} }  
		\vspace{-2ex} 
	\end{minipage} 
	\caption{Training ResNet50 on ImageNet.}
	\label{fig:loss-imagenet}
	\vspace{-1ex}
\end{figure}

Fig.~\ref{fig:loss-cifar100} and Fig.~\ref{fig:loss-imagenet} show the results of the former two settings. 
We can observe that \ourtech{} exhibits a similar convergence rate (i.e., converging efficiency in terms of epochs/iterations) comparing to the other approaches (from Fig.~\ref{fig:loss-cifar100:a} and Fig.~\ref{fig:loss-imagenet:a}), and it converges much faster than the others approaches in terms of training time (from Fig.~\ref{fig:loss-cifar100:b} and Fig.~\ref{fig:loss-imagenet:b}).
%
\ifthenelse{\boolean{isfullpaper}}
{
As the experimental results of training on MNIST, Tiny-ImageNet, and CIFAR10 are similar to the results of CIFAR100 and ImageNet shown in \autoref{fig:loss-imagenet} and \autoref{fig:loss-cifar100}, we defer them to Appendix \ref{apendex:e} for better readability. 
}
{
Besides, the experimental results of training on MNIST, Tiny-ImageNet, and CIFAR10 are similar to those shown in \autoref{fig:loss-cifar100} and \autoref{fig:loss-imagenet}, we present them in Appendix F of our \fullpaperterm{} due to the space limitation.
}

\autoref{tab:acc-nonuniform-data} summarizes the test accuracy of all the approaches with non-uniform data distribution. We observe that the test accuracy of MobileNet trained on MNIST is around 93\%, which is much lower than the typical MNIST accuracy of around 99\%. This accuracy loss is caused by the non-IID data distribution. 

Comparing to the other approaches, \ourtech{} achieves a high training speed and a comparable accuracy. 
There are two main reasons. 
First, a worker node $i$ can obtain the model from a low-link-speed neighbor $j$ in two ways. 
One is pulling the model of node $j$ directly with a low selection probability. 
The other way is pulling the models from other neighbors, by which the model information of node $j$ can be propagated to node $i$.
Second, as described in \autoref{alg:nadsgd} (lines \ref{code:l11}-\ref{code:l13}), a worker node assigns a higher weight to the pulled models from neighbors with lower selection probabilities for updating its own model. 
This design helps the worker nodes maintain enough information from the low-speed neighbors. 

\begin{table}[tp]
	\centering
	\caption{Distribution of MNIST across worker nodes}
	\begin{tabular}{l|l|l|l|l|l}
		\hline
		Worker&Lost labels&Server&Worker&Lost labels&Server\\
		\hline
		$w0$&0, 1, 2&server 1&$w4$&5,6,7&server 2 \\
		$w1$&0, 1, 3&server 1&$w5$&5,6,8&server 2 \\
	    $w2$&0, 1, 4&server 1&$w6$&5,6,9&server 2 \\
		$w3$&0, 1, 5&server 1&$w7$&5,6,0&server 2 \\
		\hline
	\end{tabular}
	\label{tab:lost-labels}
	\vspace{-1ex}
\end{table}

\begin{table}[tp]\scriptsize
	\newcommand{\multirow}
	\centering
	\caption{Accuracy of models trained over a heterogeneous network with non-uniform data partitioning}
	\begin{tabular}{ll|l|l|l|l}
		\hline
		Dataset&Model&\prague{}&Allreduce&\adpsgd{}&\ourtech{}\\
		\hline
		CIAR10&ResNet18&89.16\%&89.38\%&89.58\%&89.63\%\\	
	    CIAR100&ResNet18&71.87\%&71.28\%&71.88\%&72.17\%\\
	    MNIST&MobileNet&92.29\%&91.58\%&91.41\%&93.36\%\\
		Tiny-ImageNet&ResNet18&56.84\%&57.02\%&56.15\%&57.42\%\\
        ImageNet&ResNet50&72.37\%&73.64\%&73.08\%&73.27\%\\
		\hline
	\end{tabular}
	\label{tab:acc-nonuniform-data}
	\vspace{-1ex}
\end{table}

\subsection{Training Small Models on Complex Datasets}
\label{subsec:smallmodel}

To evaluate the effectiveness of \ourtech{} for training a small model on a complex dataset, we train MobileNet on CIFAR100. MobileNet is a small model with approximately 4.2M parameters, designed for the mobile and embedded platforms. In this set of experiments, we use the same setting as for training ResNet18 on CIFAR100 in \autoref{subsec:nonuniform-data}. In addition, we include two parameter server (PS) implementations (synchronous and asynchronous) as other baselines. For both implementations, the PS is assigned to one GPU server.

As shown in \autoref{fig:loss-ps}, \ourtech{} achieves comparable convergence rate to Prague, Allreduce-SGD, AD-PSGD, and PS with synchronous training. Furthermore, it converges much faster \textit{w.r.t.} the training time than the other competitors. 
From Fig.~\ref{fig:loss-ps:a}, PS with asynchronous training achieves the worst convergence rate. The reason is that the worker nodes located on the same server with the PS iterate much faster than the other nodes. The model maintained by the PS enhances the information from the faster nodes and weakens the information from the slower nodes, leading to a low convergence rate. Fig.~\ref{fig:loss-ps:b} shows the training loss over time. We can see that the training speed of PS with synchronous training is the worst, while the training speed of PS with asynchronous training is similar to Allreduce-SGD. 
This is mainly because of the high communication cost between the PS and the worker nodes.

\begin{table}[tp]
    \scriptsize
	\centering
	\caption{Accuracy of MobileNet trained on CIFAR100 with non-uniform data partitioning}
	\begin{tabular}{l|l|l|l|l|l}
		\hline
		\prague{} & Allreduce & \adpsgd{} & PS-syn & PS-asyn & \ourtech{} \\
		\hline
		63.32\% & 63.83\% & 63.72\% & 63.40\% & 63.77\% & 64.28\% \\	
		\hline
	\end{tabular}
	\label{tab:acc-mobilenet-ps}
	\vspace{-1ex}
\end{table}

\begin{figure}[tp]
	\centering 
	\begin{minipage}[b]{0.23\textwidth}
		\centering
		\subfigure[Loss varies with epochs] { \label{fig:loss-ps:a}     \includegraphics[width=1\textwidth]{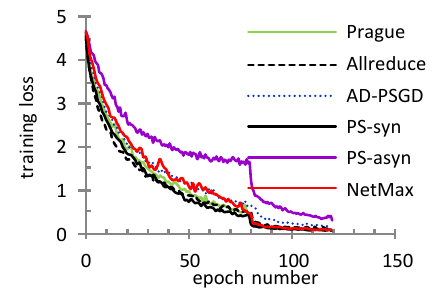} }
		\vspace{-2ex}
	\end{minipage}
	\hspace{0.01in}
	\begin{minipage}[b]{0.23\textwidth}
		\centering 
		\subfigure[Loss varies with time] { \label{fig:loss-ps:b}     \includegraphics[width=1\textwidth]{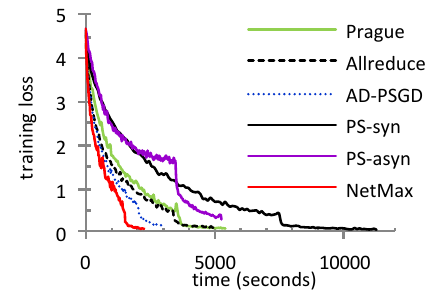} }  
		\vspace{-2ex} 
	\end{minipage} 
	\caption{Training MobileNet on CIFAR100.}
	\label{fig:loss-ps}
	\vspace{-1ex}
\end{figure}
\begin{figure}[!t]
	\centering 
	\begin{minipage}[b]{0.23\textwidth}
		\centering
		\subfigure[Loss varies with epochs] { \label{fig:loss-extend:a}     \includegraphics[width=1\textwidth]{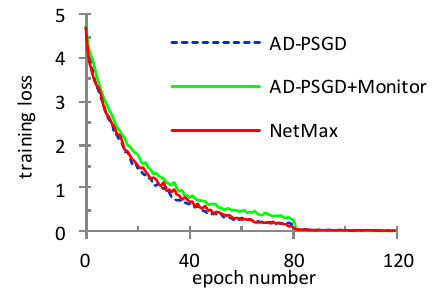} }
		\vspace{-2ex}
	\end{minipage}
	\hspace{0.01in}
	\begin{minipage}[b]{0.23\textwidth}
		\centering 
		\subfigure[Loss varies with time] { \label{fig:loss-extend:b}     \includegraphics[width=1\textwidth]{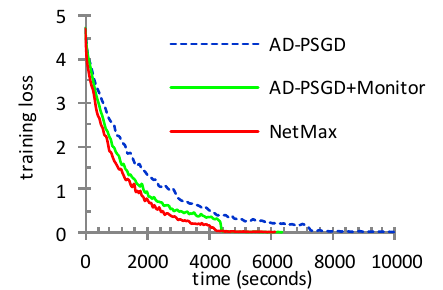} 
		}
		\vspace{-2ex} 
	\end{minipage} 
	\caption{Extension of AD-PSGD with Network Monitor}
	\label{fig:loss-extend}
	\vspace{-2ex}
\end{figure}

\autoref{tab:acc-mobilenet-ps} shows the test accuracy of MobileNet trained with various approaches. \ourtech{} achieves a similar accuracy comparing to the other competitors. 
In addition, by comparing \autoref{tab:acc-nonuniform-data} and \autoref{tab:acc-mobilenet-ps}, we observe that the accuracy of MobileNet on CIFAR100 is about 63\%, which is smaller than the 71\% accuracy of ResNet18 on CIFAR100. The reason is that MobileNet is very simple, and its capacity to learn on complex data is not as good as larger models.

\subsection{Extension of AD-PSGD with Network Monitor}
\label{subsec:extend-netmax}
We implement AD-PSGD with a Network Monitor similar to the one used in our \ourtech{} and evaluate its performance with the same setting described in \autoref{subsec:nonuniform-data} for training ResNet18 on CIFAR100.
As shown \autoref{fig:loss-extend}, AD-PSGD with a Network Monitor exhibits lower training time than the standard AD-PSGD. However, its convergence rate is lower than that of the standard AD-PSGD. The reason is similar to the one presented in \autoref{subsec:probability-performance}.
Besides, the convergence rate of AD-PSGD with a Network Monitor is slightly lower compared to \ourtech{}. This is because \ourtech{} assigns a higher weight to the model pulled from a lower-speed neighbor, while AD-PSGD uses the same weight.
Consequently, the worker nodes in AD-PSGD need more epochs to get the model information from low-speed neighbors.
}

\vspace{2ex}
\section{Related Work}
\label{sec:relwork}

\noindent\textbf{Centralized Parallel SGD.} 
The most popular implementation of the Centralized Parallel SGD (C-PSGD) is the parameter server architecture, where the models of distributed worker nodes are synchronized by a central parameter server~\cite{chilimbi2014project, chen2019round}. 
All the worker nodes exchange information with the parameter server to synchronize their training. 
The training is constrained by the network capacity at the parameter server.
\\
\textbf{Decentralized Parallel SGD.} 
To address the communication bottleneck of the central server, a number of Decentralized Parallel SGD (D-PSGD) approaches are proposed. 
The worker nodes in D-PSGD are assumed to be connected to each other such that they can synchronize their models via peer-to-peer communication. 
In this way, the traffic among worker nodes can be evenly distributed. 
Synchronous D-PSGD approaches~\cite{tang2018d,xin2019decentralized,jia2018highly} keep worker nodes training and communicating at the same pace. 
In contrast, asynchronous D-PSGD approaches~\cite{lian2018asynchronous,hegedHus2019gossip} enable worker nodes to train their local models and to communicate independently. 
Although these decentralized approaches address the problem of the centralized bottleneck, they still suffer from frequent communication via low-speed links in heterogeneous networks.
\\
\textbf{Heterogeneity-aware Distributed Training.} 
To better adapt to the heterogeneous resources, several approaches are proposed for decentralized training~\cite{luo2020prague, tang2020communication, hong2019dlion}. 
\prague{}~\cite{luo2020prague} randomly divides the worker nodes into multiple groups at each iteration such that the worker nodes in each group conduct ring-allreduce collective operation to average their models. 
The model averaging operations of different groups are independent, which can tolerate random worker slowdown. 
However, in heterogeneous networks, the node groups may still frequently communicate through low-speed links, resulting in a high communication cost. 
SAPS-PSGD~\cite{tang2020communication} assumes that the network is static and lets the worker nodes communicate with each other in a fixed topology consisting of initially high-speed links. 
However, in dynamic networks, some links of the topology used by SAPS-PSGD may become low-speed links during the training, leading to a high communication cost. 
DLion~\cite{hong2019dlion} adjusts the size of the transferred model partition according to the network link capacity. 
However, exchanging only part of the model may cause divergence of the training. 
Hop~\cite{luo2019hop} and Gaia~\cite{hsieh2017gaia} use bounded staleness for distributed settings to guarantee convergence. 
But when network links experience a continuous slowdown, the whole system would be dragged down by these low-speed links. 

In contrast, \ourtech{} enables the worker nodes to communicate efficiently via high-speed links to achieve low communication time even in dynamic networks, while guaranteeing model convergence.
\section{Conclusion}
\label{sec:conclusion}

In this paper, we propose \ourtech{}, a communication-efficient asynchronous decentralized PSGD approach, to accelerate distributed machine learning over heterogeneous networks. Compared to C-PSGD, \ourtech{} adopts decentralized training to eliminate the communication bottleneck at the central node. Compared to prior decentralized approaches, \ourtech{} can adapt to the network dynamics and enable worker nodes to communicate preferably via high-speed links. This can significantly speed up the training. In addition, we formally prove the convergence of our proposed approach.
\section*{Acknowledgments}
\label{sec:acknowledge}

The work of Qian Lin, Dumitrel Loghin, Beng Chin Ooi, and Yuncheng Wu was supported by Singapore Ministry of Education Academic Research Fund Tier 3 under MOEs official grant number MOE2017-T3-1-007. The work of Pan Zhou was supported by the China Scholarship Council.
\ifCLASSOPTIONcaptionsoff
\newpage
\fi

\bibliographystyle{IEEEtran}
\bibliography{bibliography}

\ifthenelse{\boolean{isfullpaper}}
{
\appendix

In this Appendix, we first derive the feasible intervals of $\rho$ and $\overline{t}$ used in \autoref{alg:get-policy}, followed by the bound of the approximation ratio of the feasible communication policy. Next, we prove all three theorems formulated in Section~\ref{sec:convergence-analysis}. We end by presenting additional results when (i) training models on non-uniform data partitioning and (ii) training across multiple cloud regions.

\subsection{Feasible Intervals of $\rho$ and $\overline{t}$}
\label{apendex:a}

The lower bound $L_{\rho}$ of $\rho$ is set to 0 as $\rho$ is non-negative. According to Eq. (\ref{eq:o3}), $\alpha\rho(d_{i,m}+d_{m,i}) \leq p_{i,m}\leq 1$. If worker $i$ and $m$ are neighbors, we have $\rho \leq \frac{1}{(d_{i,m}+d_{m,i})\alpha}=0.5/\alpha$.
Thus, the upper bound $U_{\rho}$ is set to $0.5/\alpha$.

By combining the constraints in Eq. (\ref{eq:o2}) and (\ref{eq:o3}), we have
\begin{equation}
\overline{t}\geq \frac{\alpha\rho}{M}\sum_{m=1}^{M}t_{i,m}(d_{i,m}+d_{m,i})=L_i, \forall i\in [M] \label{eq:oa2}
\end{equation}
Then the lower bound $L$ of $\overline{t}$ is the largest $L_i$ as follows
\begin{equation}
\overline{t}\geq L =\max_{i\in M} L_i =\max_{i\in [M]}\frac{\alpha\rho}{M}\sum_{m=1}^{M}t_{i,m}(d_{i,m}+d_{m,i}) \label{eq:oa3}
\end{equation}
For any worker node $i$, any iteration time is no larger than the iteration time when communicating with the slowest neighbor. From Eq. (\ref{eq:o2}), we have
\begin{flalign}
\overline{t}&\leq \frac{1}{M}\sum_{m=1}^{M}p_{i,m}\max_{m\in[M]}t_{i,m}d_{i,m}&\nonumber\\
&=\frac{1}{M}\max_{m\in[M]}t_{i,m}d_{i,m}=U_i, \forall i\in [M] \label{eq:oa4}&
\end{flalign}
Then the upper bound $U$ of $\overline{t}$ is the smallest $U_i$,
\begin{equation} 
\overline{t}\leq U=\min_{i\in [M]} U_i = \min_{i\in[M]}\frac{1}{M}\max_{m\in[M]}t_{i,m}d_{i,m}. \label{eq:oa5}
\end{equation}

{
\ifthenelse{\boolean{isfullpaper}}
{}
{
\color{blue}
}
\subsection{Approximation Ratio of the Feasible Communication Policy} 
\label{apendex:a-1}
We assume that the number $M$ of worker nodes is more than 3, the network connecting worker nodes is heterogeneous and fully-connected. From Eq. (\ref{eq:d2}), the matrix $\boldsymbol{Y_P}$ can be rewritten as
\begin{flalign}
\boldsymbol{Y_P} &= (1-4\alpha\rho)\boldsymbol{I} +\frac{4\alpha\rho}{M}\boldsymbol{1}\boldsymbol{1}^{T} + \boldsymbol{W} &\label{eq:a-1-1}
\end{flalign}

where the entries of $\boldsymbol{W}$ can be written as
\begin{equation}
\left\{
\begin{array}{lr}
w_{i,i}=\alpha^{2}\rho^{2}\sum_{\substack{m\in [M]\\ m\neq i}}(p_{i}p_{i,m}\gamma_{i,m}^{2}\!+\!p_{m}p_{m,i}\gamma_{m,i}^{2}), \forall i\in [M]& \\
&\\  
w_{i,m} = -\alpha^{2}\rho^{2}(p_{i}p_{i,m}\gamma_{i,m}^{2}\!+\!p_{m}p_{m,i}\gamma_{m,i}^{2}),\forall i,m\!\in\! [M],m\!\neq\! i & 
\end{array}
\right. \label{eq:a-1-2}
\end{equation} 
From Eq. (\ref{eq:a-1-2}), the sum of the rows of $\boldsymbol{W}$ is 0. Therefore, $\lambda_{\boldsymbol{W}}=0$ is one of the eigenvalues of $\boldsymbol{W}$. In addition, as $Tr(\boldsymbol{W})$ is positive, $\lambda_{\boldsymbol{W}}=0$ is not the largest eigenvalue $\lambda_{\boldsymbol{W},1}$ of $\boldsymbol{W}$. Let $\boldsymbol{Z}=(1-4\alpha\rho)\boldsymbol{I} +\frac{4\alpha\rho}{M}\boldsymbol{1}\boldsymbol{1}^{T}$, then the eigenvalues $\lambda_{\boldsymbol{Z},1}\geq \lambda_{\boldsymbol{Z},2}\geq\dots\geq\lambda_{\boldsymbol{Z},M}$ of $\boldsymbol{Z}$ are $1,1\!-\!4\alpha\rho,\dots,1\!-\!4\alpha\rho$. 

According to~\cite{1999Eigenvalues}, there exists the following relationship among the eigenvalues of $\boldsymbol{Y_P}$, $\boldsymbol{W}$ and $\boldsymbol{Z}$,
\begin{equation} 
\lambda_{M-i-j} \geq \lambda_{\boldsymbol{Z},M-i} + \lambda_{\boldsymbol{W},M-j} \label{eq:a-1-3}
\end{equation}

Let $i=0$ and $\lambda_{\boldsymbol{W},M-j} = 0$, then Eq. (\ref{eq:a-1-3}) becomes
\begin{equation} 
\lambda_2 \geq \lambda_{\boldsymbol{Z},M}\geq 1-4\alpha\rho \label{eq:a-1-4}
\end{equation}

When the graph connecting the worker nodes is fully-connected, from Eq. (\ref{eq:o3}), we have that $p_{i,m}\geq 2\alpha\rho, \forall i,m \in [M], i\neq m$. As $\sum_{m\in [M],m\neq i}p_{i,m}\leq 1, \forall i\in[M]$, we have
\begin{equation} 
\alpha\rho \leq \frac{1}{2(M-1)}\label{eq:a-1-5}
\end{equation}
By using Eq. (\ref{eq:a-1-4}) and Eq. (\ref{eq:a-1-5}), we obtain the lower bound of the second largest eigenvalue of $\boldsymbol{Y_P}$ as follows
\begin{equation} 
\lambda_2\geq \frac{M-3}{M-1}\label{eq:a-1-6}
\end{equation}

Let $a$ be the minimum positive entry in
$\boldsymbol{Y_P}$. Based on~\cite{Kirkland2009A}, the second largest eigenvalue $\lambda_2$ of $\boldsymbol{Y_P}$ can be bounded by
\begin{equation} 
\lambda_2 \leq \frac{1-2a+a^{M+1}}{1-2a+a^{M}}\label{eq:a-1-7}
\end{equation}

Let $\lambda^*$ be the second largest eigenvalue when the objective function in Eq. (\ref{eq:o0}) reaches the optimal $l(\lambda^*)$. By combining Eq. (\ref{eq:o1}) and Eq. (\ref{eq:a-1-6}), we have that
\begin{equation} 
l(\lambda^*)=\overline{t}\frac{\ln\varepsilon}{\ln\lambda^*}\geq \overline{t}\frac{\ln\varepsilon}{\ln\frac{M-3}{M-1}}\geq L\frac{\ln\varepsilon}{\ln\frac{M-3}{M-1}} \label{eq:a-1-8}
\end{equation}

Let $l(\lambda_2)$ denote the objective function of the feasible communication policy derived by \autoref{alg:get-policy}. By combining Eq. (\ref{eq:o1}) and Eq. (\ref{eq:a-1-7}), we have that
\begin{equation} 
l(\lambda_2)=\overline{t}\frac{\ln\varepsilon}{\ln\lambda_2}\leq U\frac{\ln\varepsilon}{\ln\frac{1-2a+a^{M+1}}{1-2a+a^{M}}} \label{eq:a-1-9}
\end{equation}

By combining Eq. (\ref{eq:a-1-8}) and Eq. (\ref{eq:a-1-9}), we derive the bound of the approximation ratio as
\begin{flalign}
\frac{l(\lambda_2)}{l(\lambda^*)} \leq \frac{U}{L}\frac{\ln(M\!-\!1)\!-\!\ln(M\!-\!3)}{\ln(1\!-\!2a\!+\!a^{M})\!-\!\ln(1\!-\!2a\!+\!a^{M\!+\!1})}
\label{eq:a-1-10}&
\end{flalign}
}

\subsection{Proof of Theorem~\ref{theorem:static}} \label{apendex:b}

\emph{Proof.} For simplicity, $\nabla f(x_{i}^{k};\mathcal{D}_i)$ denotes $\nabla f(x_{i}^{k})$. Let $\lambda_{1}$ and $\lambda_{2}$ be the largest and second largest eigenvalue of matrix $\mathbb{E}[(\boldsymbol{D}^k)^{T}\boldsymbol{D}^k]$, respectively.
We can bound the following expected sum of squares deviation by applying Eq. (\ref{eq:c3}).
\begin{flalign}
&\mathbb{E}\Big[ \lVert \boldsymbol{x}^{k+1}-x^{*}\boldsymbol{1}\rVert^{2}|\boldsymbol{x}^k\Big] &\nonumber\\ 
&= \mathbb{E}\Big[\lVert \boldsymbol{D}^{k}(\boldsymbol{x}^{k}-\alpha \boldsymbol{g}^{k})-x^{*}\boldsymbol{1}\rVert^{2} |\boldsymbol{x}^k\Big]&\nonumber\\
&= \mathbb{E}\Big[ (\boldsymbol{x}^{k}-x^{*}\boldsymbol{1}-\alpha \boldsymbol{g}^k)^{T}(\boldsymbol{D}^k)^T\boldsymbol{D}^k (\boldsymbol{x}^k-x^{*}\boldsymbol{1}-\alpha \boldsymbol{g}^k)|\boldsymbol{x}^k \Big]&\nonumber\\
&\leq \lambda\mathbb{E}\Big[ (\boldsymbol{x}^k-x^{*}\boldsymbol{1}-\alpha \boldsymbol{g}^k)^{T}(\boldsymbol{x}^k-x^{*}\boldsymbol{1}-\alpha \boldsymbol{g}^k)|\boldsymbol{x}^k \Big]&\label{eq:c5}
\end{flalign}
where $\lambda=\lambda_{1}$. If $\mathbb{E}[(\boldsymbol{D}^{k})^{T}\boldsymbol{D}^{k}]$ is doubly stochastic matrix, $\lambda = \lambda_{2}$ \cite{boyd2006randomized,lian2018asynchronous}. 

\begin{flalign}
&\mathbb{E}\Big[ (\boldsymbol{x}^k-x^{*}\boldsymbol{1}-\alpha \boldsymbol{g}^k)^{T}(\boldsymbol{x}^k-x^{*}\boldsymbol{1}-\alpha \boldsymbol{g}^k)|\boldsymbol{x}^k \Big]&\nonumber\\
&=\mathbb{E}\Big[ \lVert \boldsymbol{x}^k-x^{*}\boldsymbol{1}\rVert^{2}\!-\!2\alpha (\boldsymbol{g}^k)^{T}(\boldsymbol{x}^k\!-\!x^{*}\boldsymbol{1})+\alpha^{2}(\boldsymbol{g}^{k})^{T}\boldsymbol{g}^k |\boldsymbol{x}^k\Big]&\nonumber\\
&=\lVert \boldsymbol{x}^k-x^{*}\boldsymbol{1}\rVert^{2} - 2\alpha \sum_{n=1}^{M} p_{n}\nabla f(x_{n}^{k})^{T}(x_{n}^{k}-x^{*}) &\nonumber\\ 
&\quad+\! \alpha^{2} \!\sum_{n=1}^{M} p_{n}\nabla f(x_{n}^{k})^{T}\nabla f(x_{n}^{k}) \!+\! \alpha^{2}\!\sum_{n}^{M}p_{n}\mathbb{E}\Big[(\xi_{n}^{k})^T\xi_{n}^{k}\Big]& \label{eq:c7}
\end{flalign}

In Eq. (\ref{eq:c7}), we drop the terms that are linear in $\xi_{i}^{k}$ since $\xi_{i}^{k}$ has zero mean assumption. To bound the second term at the right side of Eq. (\ref{eq:c7}), we apply the property of convex functions, 
\begin{flalign}
&(\nabla f(x)-\nabla f(z))^T (x-z)&\nonumber\\
&\geq \frac{\mu L}{\mu +L}(x-z)^T(x-z)&\nonumber\\
&\quad+\!\frac{1}{\mu\!+\!L}(\nabla f(x)\!-\!\nabla f(z))^T(\nabla f(x)\!-\!\nabla f(z)), \forall x,z& \label{eq:c8}
\end{flalign}
With $x=x_{i}^{k}$ and $z=x^{*}$, we have that\\
\begin{flalign}
& -\nabla f(x_{n}^{k})^{T}(x_{n}^{k}-x^{*}) = -\nabla (f(x_{n}^{k})-0)^{T}(x_{n}^{k}-x^{*}) &\nonumber\\
&\!\!\leq -\frac{\mu L}{\mu\!\! +\!\!L}(x_{n}^{k}\!-\!x^*)^T(x_{n}^{k}\!-\!x^*)\!-\!\frac{1}{\mu\!\!+\!\!L}\nabla\! f(x_{n}^{k})^T\nabla\! f(x_{n}^{k})& \label{eq:c9}
\end{flalign}
By applying Eq. (\ref{eq:c9}), we can bound
\begin{flalign}
& - 2\alpha \sum_{n=1}^{M} p_{n}\nabla f(x_{n}^{k})^{T}(x_{n}^{k}-x^{*})&\nonumber\\
&\leq -\frac{2\alpha\mu L}{\mu +L}\sum_{n=1}^{M}p_{n}(x_{n}^{k}-x^*)^T(x_{n}^{k}-x^*)&\nonumber\\
&\quad-\frac{2\alpha}{\mu +L}\sum_{n=1}^{M}p_{n}\nabla f(x_{n}^{k})^T\nabla f(x_{n}^{k})& \nonumber\\
&\leq -\frac{2\alpha\mu L}{\mu +L}p_{min}\sum_{n=1}^{M}(x_{n}^{k}-x^*)^T(x_{n}^{k}-x^*)&\nonumber\\
&\quad-\frac{2\alpha}{\mu +L}\sum_{n=1}^{M}p_{n}\nabla f(x_{n}^{k})^T\nabla f(x_{n}^{k})& \label{eq:c10}
\end{flalign}
where $p_{min}$ is the samllest value among $p_{n},\forall n \!\!\in\!\![M]$. Combining Eq.(\ref{eq:c7}) and Eq. (\ref{eq:c10}), we have that
\begin{flalign}
&\mathbb{E}\Big[ (\boldsymbol{x}^k-x^{*}\boldsymbol{1}-\alpha \boldsymbol{g}^k)^{T}(\boldsymbol{x}^k-x^{*}\boldsymbol{1}-\alpha \boldsymbol{g}^k)|\boldsymbol{x}^k \Big]&\nonumber\\
&\leq (1-\frac{2\alpha\mu L}{\mu +L}p_{min})\lVert \boldsymbol{x}^k-x^{*}\boldsymbol{1}\rVert^{2}+\alpha^{2}\sigma^{2}&\label{eq:c12}\\
&\quad+(\alpha^{2}-\frac{2\alpha}{\mu +L})\sum_{n=1}^{M}p_{n}\nabla f(x_{n}^{k})^T\nabla f(x_{n}^{k})& \label{eq:c11}
\end{flalign}
The term in Eq. (\ref{eq:c11}) can be dropped if $0<\alpha<\frac{2}{\mu+L}$. Thus, we have 
\begin{flalign}
&\mathbb{E}\Big[ (\boldsymbol{x}^k-x^{*}\boldsymbol{1}-\alpha \boldsymbol{g}^{k})^{T}(\boldsymbol{x}^k-x^{*}\boldsymbol{1}-\alpha \boldsymbol{g}^k)|\boldsymbol{x}^k \Big]&\nonumber\\
&\leq (1-\frac{2\alpha\mu L}{\mu +L}p_{min})\lVert \boldsymbol{x}^k-x^{*}\boldsymbol{1}\rVert^{2}+\alpha^{2}\sigma^{2}& \label{eq:c13}
\end{flalign}
By applying Eq. (\ref{eq:c13}) in Eq (\ref{eq:c5}), we have that
\begin{flalign}
&\mathbb{E}\Big[ \lVert \boldsymbol{x}^{k+1}-x^{*}\boldsymbol{1}\rVert^{2}|\boldsymbol{x}^k\Big]&\nonumber \\ 
&\leq \lambda\mathbb{E}\Big[ (\boldsymbol{x}^k-x^{*}\boldsymbol{1}-\alpha \boldsymbol{g}^k)^{T}(\boldsymbol{x}^k-x^{*}\boldsymbol{1}-\alpha \boldsymbol{g}^k)|\boldsymbol{x}^k \Big]&\nonumber \\
&\leq \lambda(1-\frac{2\alpha\mu L}{\mu +L}p_{min})\lVert \boldsymbol{x}^k-x^{*}\boldsymbol{1}\rVert^{2}+ \lambda\alpha^{2}\sigma^{2}&\nonumber \\
&\leq \lambda\lVert \boldsymbol{x}^k-x^{*}\boldsymbol{1}\rVert^{2}+ \lambda\alpha^{2}\sigma^{2}&\label{eq:c14}
\end{flalign}
By unrolling the recursion, we have that
\begin{flalign}
\mathbb{E}\Big[ \lVert \boldsymbol{x}^k-x^{*}\boldsymbol{1}\rVert^{2}\Big] 
&\leq \lambda^{k}\lVert \boldsymbol{x}^{0}-x^{*}\boldsymbol{1}\rVert^{2}+ \frac{1-\lambda^{k}}{1-\lambda}\lambda\alpha^{2}\sigma^{2}&\nonumber\\ 
&\leq \lambda^{k}\lVert \boldsymbol{x}^{0}-x^{*}\boldsymbol{1}\rVert^{2}+ \frac{\lambda}{1-\lambda}\alpha^{2}\sigma^{2}.\Box &\label{eq:c15}
\end{flalign}

\subsection{Proof of Theorem~\ref{theorem:dynamic}}
\label{apendex:c}

\emph{Proof.} For the convergence analysis under dynamic networks, we assume that the network changes at each step $k$. Thus, we consider different values of $\lambda$ at each step, denoted by $\lambda_k$. Then, we have
\begin{equation}
\mathbb{E}\Big[ \lVert \boldsymbol{x}^k-x^{*}\boldsymbol{1}\rVert^{2}|\boldsymbol{x}^{k-1}\Big] \leq \lambda_k \lVert \boldsymbol{x}^{k-1}-x^{*}\boldsymbol{1} \rVert^{2}+ \alpha^{2}\sigma^{2}\lambda_k \label{eq:h0}
\end{equation}
By unrolling the recursion, we have that
\begin{flalign}
&\mathbb{E}\Big[ \lVert \boldsymbol{x}^k-x^{*}\boldsymbol{1}\rVert^{2}\Big]&\nonumber\\ 
&\leq \Big[\prod_{n=1}^{k}\lambda_n\Big]\lVert \boldsymbol{x}^{0}-x^{*}\boldsymbol{1}\rVert^{2} + \sum_{n=1}^{k}\lambda_n\alpha^{2}\sigma^{2}\prod_{m=n+1}^{k}\lambda_{max}& \nonumber\\
&\leq \lambda_{max}^{k}\lVert \boldsymbol{x}^{0}-x^{*}\boldsymbol{1}\rVert^{2} + \sum_{n=1}^{k}\lambda_{max}\alpha^{2}\sigma^{2}\lambda_{max}^{k-n}& \nonumber\\
&\leq \lambda_{max}^k\lVert \boldsymbol{x}^{0}-x^{*}\boldsymbol{1}\rVert^{2} + \frac{1-\lambda_{max}^k}{1-\lambda_{max}}\lambda_{max}\alpha^{2}\sigma^{2}&\nonumber\\
&\leq \lambda_{max}^k\lVert \boldsymbol{x}^{0}-x^{*}\boldsymbol{1}\rVert^{2} + \frac{\lambda_{max}}{1-\lambda_{max}}\alpha^{2}\sigma^{2}. \Box &\label{eq:h1}
\end{flalign}

\subsection{Proof of Theorem~\ref{theorem:converge}} \label{apendex:d}

Before proving Theorem~\ref{theorem:converge}, we first introduce three lemmas.

\begin{lemma}
\label{lemma:1}
If the probability matrix $\boldsymbol{P}$ is a feasible solution of the optimization problem in Eq. (\ref{eq:o0})-(\ref{eq:o5}), then $\boldsymbol{Y}_{\boldsymbol{P}}$ is a symmetric matrix and each row/column of $\boldsymbol{Y}_{\boldsymbol{P}}$ sums to 1. 
\end{lemma}

\emph{Proof.} As $\boldsymbol{P}$ is a feasible solution, the constraints in Eq. (\ref{eq:o2}) are satisfied and we have $\overline{t}_{i} = \sum_{m=1}^{M}p_{i,m}t_{i,m}d_{i,m} = M\overline{t},\forall i \in [M]$.
Using Eq. (\ref{eq:d4}), we have
$
p_{i} = \frac{1/\overline{t}_{i}}{\sum_{m=1}^{M}1/\overline{t}_{m}} = \frac{1}{M},\forall i \in [M]
$.
Using Eq. (\ref{eq:d2}), the rows of $\boldsymbol{Y}_{\boldsymbol{P}}$ sum to 
\begin{flalign}
\sum_{m=1}^{M}y_{i,m} &=1-\alpha\rho\!\!\sum_{\substack{m\in [M]\\ m\neq i}}\!p_{i}p_{i,m}\gamma_{i,m}+\alpha\rho\!\!\sum_{\substack{m\in [M]\\ m\neq i}}\!p_{m}p_{m,i}\gamma_{i,m}\nonumber&\\
&= 1-\alpha\rho\!\!\sum_{\substack{m\in [M]\\ m\neq i}}\!\!\frac{1}{M}(p_{i,m}\gamma_{i,m}-p_{m,i}\gamma_{m,i}) \nonumber&\\
&= 1, \forall i \in [M]& \label{eq:lemma1-03}
\end{flalign}
From Eq. (\ref{eq:d2}), $y_{i,m}=y_{m,i}$, thus $\boldsymbol{Y}_{\boldsymbol{P}}$ is symmetric and the columns of $\boldsymbol{Y}_{\boldsymbol{P}}$ also sum to 1. $\Box$

\begin{lemma}
\label{lemma:2}
If the probability matrix $\boldsymbol{P}$ is a feasible solution of optimization problem in Eq. (\ref{eq:o0})-(\ref{eq:o5}), then $\boldsymbol{Y}_{\boldsymbol{P}}$ is non-negative.
\end{lemma}

\emph{Proof.} As $\boldsymbol{P}$ is a feasible solution, the constraints in Eq. (\ref{eq:o3}) are satisfied. By applying Eq. (\ref{eq:o3}), for any $p_{i,m}\neq0, i\neq m$, we have that
\begin{equation}
\alpha\rho\gamma_{i,m}=\frac{\alpha\rho(d_{i,m}+d_{m,i})}{p_{i,m}}<1 \label{eq:lemma2-01}
\end{equation}
By applying Eq. (\ref{eq:d2}) and (\ref{eq:lemma2-01}), for any $p_{i,m}\neq0, i\neq m$, we have that
\begin{flalign}
y_{i,m} &=\alpha\rho p_{i}p_{i,m}\gamma_{i,m}(1\!-\!\alpha\rho\gamma_{i,m})\!+\!\alpha\rho p_{m}p_{m,i}\gamma_{m,i}(1\!-\!\alpha\rho\gamma_{m,i})\nonumber&\\ 
&>0&\label{eq:lemma2-02}
\end{flalign}
For any $p_{i,m}=0, i\neq m$, $y_{i,m}=0$. Thus, for any $i,m \in [M], i\neq m$, we have $y_{i,m}\geq 0$.\\
Combining Eq. (\ref{eq:d2}) with $p_{i}=\frac{1}{M}$, for any $i\in [M]$, we have

\begin{flalign}
y_{i,i} &=1-\frac{2\alpha\rho}{M} \sum_{\substack{m\in [M]\\ m\neq i}}(d_{i,m}+d_{m,i})\nonumber&\\
& \qquad+ \frac{\alpha^2\rho^2}{M} \sum_{\substack{m\in [M]\\ m\neq i}}\Big[\frac{(d_{i,m}+d_{m,i})^2}{p_{i,m}}+\frac{(d_{i,m}+d_{m,i})^2}{p_{m,i}}\Big]\nonumber& \\
&\geq 1\!-\!\frac{2\alpha\rho}{M}\!\! \sum_{\substack{m\in [M]\\ m\neq i}}\!\!(d_{i,m}\!+\!d_{m,i})\!+\!\frac{2\alpha^2\rho^2}{M}\!\!\!\sum_{\substack{m\in [M]\\ m\neq i}}\!\!\!(d_{i,m}\!+\!d_{m,i})^2\nonumber&\\
&= 1- \frac{2\alpha\rho}{M}\!\! \sum_{\substack{m\in [M]\\ m\neq i}}\!\!\Big[(d_{i,m}\!+\!d_{m,i})\big(1\!-\!\alpha\rho(d_{i,m}\!+\!d_{m,i})\big)\Big]\label{eq:lemma2-03}&\\
&\geq 0& \label{eq:lemma2-04}
\end{flalign}
When $\alpha\rho > 0.5$, the term $(d_{i,m}+d_{m,i})\big(1-\alpha\rho(d_{i,m}+d_{m,i})$ in Eq. (\ref{eq:lemma2-03}) is negative. Eq. (\ref{eq:lemma2-03}) reaches the smallest value of $1-\frac{4\alpha\rho}{M}(1-2\alpha\rho)$ when worker node $i$ has only one neighbor. Since $M\geq 2$, the item $1-\frac{4\alpha\rho}{M}(1-2\alpha\rho)$ is non-negative and, thus, Eq. (\ref{eq:lemma2-03}) is non-negative. When  $\alpha\rho \leq 0.5$, the term $(d_{i,m}+d_{m,i})\big(1-\alpha\rho(d_{i,m}+d_{m,i})$ is non-negative. Eq. (\ref{eq:lemma2-03}) reaches the smallest value of $1-\frac{M-1}{M}4\alpha\rho(1-2\alpha\rho)$ when worker node $i$ and all the other nodes are neighbors. As $1-\frac{M-1}{M}4\alpha\rho(1-2\alpha\rho)$ is larger than the positive polynomial $1-4\alpha\rho(1-2\alpha\rho)$, Eq. (\ref{eq:lemma2-03}) is positive. Therefore, Eq. (\ref{eq:lemma2-04}) always holds. In summary, $\boldsymbol{Y}_{\boldsymbol{P}}$ is non-negative. $\Box$
\\

Let graph $\mathcal{\boldsymbol{G}}_{\boldsymbol{A}}$ be an $n\times n$ matrix $\boldsymbol{A}=[a_{i,m}]_{n\times n}$ with a vertex set $\mathcal{\boldsymbol{V}}=\{1,2,\dots,n\}$. There is an edge from vertex $m$ to $i$ if and only if $a_{i,m} \neq 0$. Then $\mathcal{\boldsymbol{G}}_{\boldsymbol{A}}$ is \textbf{connected} if there is a path between every pair of vertices \cite{rigo2016advanced}.

\begin{lemma}
\label{lemma:3}
Let $\boldsymbol{P}$ be a feasible solution of the optimization problem in Eq. (\ref{eq:o0})-(\ref{eq:o5}). If the graph $\mathcal{\boldsymbol{G}}_{\boldsymbol{P}}$ of $\boldsymbol{P}$ is connected, then the graph $\mathcal{\boldsymbol{G}}_{\boldsymbol{Y}}$ of $\boldsymbol{Y}_{\boldsymbol{P}}$ is also connected.
\end{lemma}
\emph{Proof.} From  Eq. (\ref{eq:lemma2-01})-(\ref{eq:lemma2-02}), we have that if $p_{i,m}\neq 0$, then $y_{i,m}\neq 0$. This means that for any path in $\mathcal{\boldsymbol{G}}_{\boldsymbol{P}}$, there is also a path in $\mathcal{\boldsymbol{G}}_{\boldsymbol{Y}}$. Thus, if the graph $\mathcal{\boldsymbol{G}}_{\boldsymbol{P}}$ is connected, then the graph $\mathcal{\boldsymbol{G}}_{\boldsymbol{Y}}$ is also connected. $\Box$
\\


\emph{Proof of Theorem~\ref{theorem:converge}.} From Lemma~\ref{lemma:2} and Lemma~\ref{lemma:3}, we obtain that $\boldsymbol{Y}_{\boldsymbol{P}}$ is non-negative and the graph $\mathcal{\boldsymbol{G}}_{\boldsymbol{Y}}$ of $\boldsymbol{Y}_{\boldsymbol{P}}$ is connected under Assumption 1, which means that $\boldsymbol{Y}_{\boldsymbol{P}}$ is irreducible~\cite{rigo2016advanced}. Meanwhile, according to the Perron-Frobenius theorem \cite{rigo2016advanced}, the non-negative irreducible $\boldsymbol{Y}_{\boldsymbol{P}}$ has an unique positive real eigenvalue equal to spectral radium. Moreover, the eigenvalues of $\boldsymbol{Y}_{\boldsymbol{P}}$ are all real since $\boldsymbol{Y}_{\boldsymbol{P}}$ is symmetric (Lemma~\ref{lemma:1}). Thus, the largest eigenvalue of $\boldsymbol{Y}_{\boldsymbol{P}}$ is equal to the spectral radium and is unique. From Lemma~\ref{lemma:1} and Lemma~\ref{lemma:2}, we derive that $\boldsymbol{Y}_{\boldsymbol{P}}$ is doubly stochastic symmetric. Thus, for any feasible solution $\boldsymbol{P}$, the largest eigenvalue of $\boldsymbol{Y}_{\boldsymbol{P}}$ is 1~\cite{gershgorin2014gershgorin, lian2018asynchronous} and the second largest eigenvalue of $\boldsymbol{Y}_{\boldsymbol{P}}$ is strictly less than 1. Since the policy $\boldsymbol{P}$ derived by \autoref{alg:get-policy}
is a feasible solution for the optimization problem in Eq. (\ref{eq:o0})-(\ref{eq:o5}), then $\boldsymbol{Y}_{\boldsymbol{P}}$ is a doubly stochastic matrix with the second largest eigenvalue strictly less than 1.
Based on Theorem~\ref{theorem:static} and Theorem~\ref{theorem:dynamic}, our decentralized training approach converges. 

{
\ifthenelse{\boolean{isfullpaper}}
{}
{
\color{blue}
}
Combining Eq. (\ref{eq:c5}) and Eq. (\ref{eq:c7}), we have that 
\begin{flalign}
&\mathbb{E}\Big[ \lVert \boldsymbol{x}^{k+1}-x^{*}\boldsymbol{1}\rVert^{2}\Big]&\nonumber \\ 
&\leq \mathbb{E}\Big[ \lVert \boldsymbol{x}^k-x^{*}\boldsymbol{1}\rVert^{2}\Big] - 2\alpha \sum_{n=1}^{M} p_{n}\mathbb{E}\Big[ \nabla f(x_{n}^{k})^{T}(x_{n}^{k}-x^{*}) \Big]&\nonumber\\ 
&\quad+\! \alpha^{2} \!\sum_{n=1}^{M} p_{n} \mathbb{E}\Big[ \nabla f(x_{n}^{k})^{T}\nabla f(x_{n}^{k})\Big] \!+\! \alpha^{2}\!\sum_{n}^{M}p_{n}\mathbb{E}\Big[(\xi_{n}^{k})^T\xi_{n}^{k}\Big]&\label{eq:theorem3-1}
\end{flalign}
By applying Assumption 1 and $p_n =\frac{1}{M}$, Eq. (\ref{eq:theorem3-1}) can be rewritten as follows
\begin{flalign}
&\mathbb{E}\Big[ \lVert \boldsymbol{x}^{k+1}-x^{*}\boldsymbol{1}\rVert^{2}\Big]&\nonumber \\ 
&\leq \mathbb{E}\Big[ \lVert \boldsymbol{x}^k-x^{*}\boldsymbol{1}\rVert^{2}\Big] - \frac{2\alpha}{M} \sum_{n=1}^{M}\mathbb{E}\Big[ \nabla f(x_{n}^{k})^{T}(x_{n}^{k}-x^{*}) \Big]&\nonumber\\ 
&\quad+\! \alpha^{2} \eta^2 + \alpha^{2}\delta^2&\label{eq:theorem3-2}
\end{flalign}
which we rearrange as
\begin{flalign}
&\frac{2\alpha}{M} \sum_{n=1}^{M}\mathbb{E}\Big[ \nabla f(x_{n}^{k})^{T}(x_{n}^{k}-x^{*}) \Big] &\nonumber\\
&\leq \mathbb{E}\Big[ \lVert \boldsymbol{x}^k\!-\!x^{*}\boldsymbol{1}\rVert^{2}\Big]\!-\! \mathbb{E}\Big[ \lVert \boldsymbol{x}^{k+1}\!-\!x^{*}\boldsymbol{1}\rVert^{2}\Big]\!+\! \alpha^{2} (\eta^2 \!+\! \delta^2)&\label{eq:theorem3-3}
\end{flalign}
From the convexity of function $f$, we have that
\begin{equation}
\nabla f(x)^T(x-x^*)\geq f(x)-f(x^*) \label{eq:theorem3-4}
\end{equation}
Combining Eq. (\ref{eq:theorem3-3}) and (\ref{eq:theorem3-4}), we have that
\begin{flalign}
&2\alpha \mathbb{E}\Big[\frac{1}{M}\sum_{n=1}^{M}\big(f(x_{n}^{k})-f(x^{*})\big) \Big] &\nonumber \\
&\leq \mathbb{E}\Big[ \lVert \boldsymbol{x}^k\!-\!x^{*}\boldsymbol{1}\rVert^{2}\Big]\!-\! \mathbb{E}\Big[ \lVert \boldsymbol{x}^{k+1}\!-\!x^{*}\boldsymbol{1}\rVert^{2}\Big]\!+\! \alpha^{2} (\eta^2 \!+\! \delta^2)&\label{eq:theorem3-5}
\end{flalign}
Summarizing the inequality above from $l=1,2,\dots,k$, we have 
\begin{flalign}
&2\alpha \sum_{l=1}^{k}\mathbb{E}\Big[\frac{1}{M}\sum_{n=1}^{M}\big(f(x_{n}^{l})-f(x^{*})\big) \Big] &\nonumber \\
&\leq \mathbb{E}\Big[ \lVert \boldsymbol{x}^1\!-\!x^{*}\boldsymbol{1}\rVert^{2}\Big]\!-\! \mathbb{E}\Big[ \lVert \boldsymbol{x}^{k+1}\!-\!x^{*}\boldsymbol{1}\rVert^{2}\Big]\!+\! k\alpha^{2} (\eta^2 \!+\! \delta^2)&\nonumber\\
&\leq \mathbb{E}\Big[ \lVert \boldsymbol{x}^1\!-\!x^{*}\boldsymbol{1}\rVert^{2}\Big]\!+\! k\alpha^{2} (\eta^2 \!+\! \delta^2)&\label{eq:theorem3-6}
\end{flalign}
Combining Eq. (\ref{eq:c15}) and (\ref{eq:theorem3-6}), we have 
\begin{flalign}
&2\alpha \sum_{l=1}^{k}\mathbb{E}\Big[\frac{1}{M}\sum_{n=1}^{M}\big(f(x_{n}^{l})-f(x^{*})\big) \Big] &\nonumber \\
&\leq \lambda \lVert \boldsymbol{x}^0\!-\!x^{*}\boldsymbol{1}\rVert^{2}\!+\! \lambda\alpha^2\delta^2+ k\alpha^{2} (\eta^2 \!+\! \delta^2)&\label{eq:theorem3-7}
\end{flalign}
which we rearrange as 
\begin{flalign}
&\frac{\sum_{l=1}^{k}\mathbb{E}\Big[\frac{1}{M}\sum_{n=1}^{M}\big(f(x_{n}^{l})-f(x^{*})\big) \Big]}{k} &\nonumber \\
&\leq \frac{\lambda \lVert \boldsymbol{x}^0\!-\!x^{*}\boldsymbol{1}\rVert^{2}\!+\! \lambda\alpha^2\delta^2+ k\alpha^{2} (\eta^2 \!+\! \delta^2)}{2k\alpha}&\label{eq:theorem3-8}
\end{flalign}
For a chosen constant $\alpha=\frac{c}{\sqrt{k}}$, the we have
\begin{flalign}
&\frac{\sum_{l=1}^{k}\mathbb{E}\Big[\frac{1}{M}\sum_{n=1}^{M}\big(f(x_{n}^{l})-f(x^{*})\big) \Big]}{k} &\nonumber \\
&\leq \frac{\lambda \lVert \boldsymbol{x}^0\!-\!x^{*}\boldsymbol{1}\rVert^{2}}{2c\sqrt{k}}\!+\! \frac{c\lambda\delta^2}{2k\sqrt{k}}+ \frac{c(\eta^2 \!+\! \delta^2)}{2\sqrt{k}}&\label{eq:theorem3-9}
\end{flalign}

The above inequality means that the decentralized training converges to the optima $x^*$ with a rate $O(1/\sqrt{k})$. This completes the proof.$\Box$

\subsection{Training Models on Non-uniform Data Partitioning} \label{apendex:e}

\begin{figure}[tp]
	\centering 
	\begin{minipage}[b]{0.23\textwidth}
		\centering
		\subfigure[Loss varies with epochs] { \label{fig:loss-cifar10:a}     \includegraphics[width=1\textwidth]{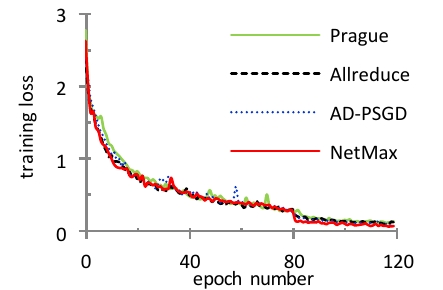} }
		\vspace{-2ex}
	\end{minipage}
	\hspace{0.01in}
	\begin{minipage}[b]{0.23\textwidth}
		\centering 
		\subfigure[Loss varies with time] { \label{fig:loss-cifar10:b}     \includegraphics[width=1\textwidth]{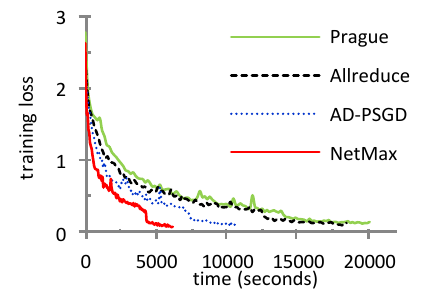} }  
		\vspace{-2ex} 
	\end{minipage} 
	\caption{Training ResNet18 on CIFAR10.}
	\label{fig:loss-cifar10}
	\vspace{-1ex}
\end{figure}

\begin{figure}[tp]
	\centering 
	\begin{minipage}[b]{0.23\textwidth}
		\centering
		\subfigure[Loss varies with epochs] { \label{fig:loss-tiny:a}     \includegraphics[width=1\textwidth]{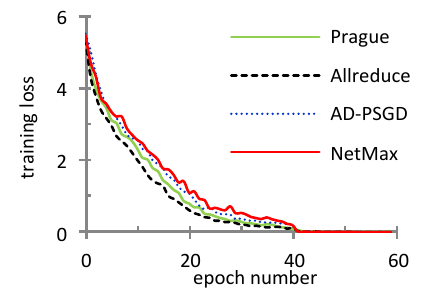} }
		\vspace{-2ex}
	\end{minipage}
	\hspace{0.01in}
	\begin{minipage}[b]{0.23\textwidth}
		\centering 
		\subfigure[Loss varies with time] { \label{fig:loss-tiny:b}     \includegraphics[width=1\textwidth]{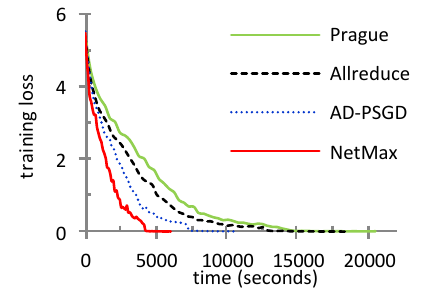} }  
		\vspace{-2ex} 
	\end{minipage} 
	\caption{Training ResNet18 on Tiny-ImageNet.}
	\label{fig:loss-tiny}
	\vspace{-1ex}
\end{figure}

\begin{figure}[!tp]
	\centering 
	\begin{minipage}[b]{0.23\textwidth}
		\centering
		\subfigure[Loss varies with iterations] { \label{fig:loss-mnist:a}     \includegraphics[width=1\textwidth]{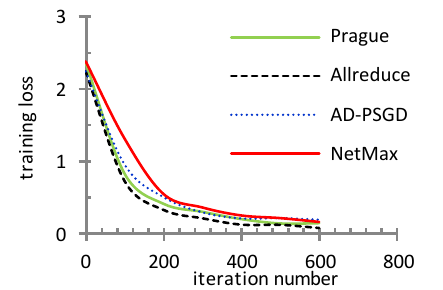} }
		\vspace{-2ex}
	\end{minipage}
	\hspace{0.01in}
	\begin{minipage}[b]{0.23\textwidth}
		\centering 
		\subfigure[Loss varies with time] { \label{fig:loss-mnist:b}     \includegraphics[width=1\textwidth]{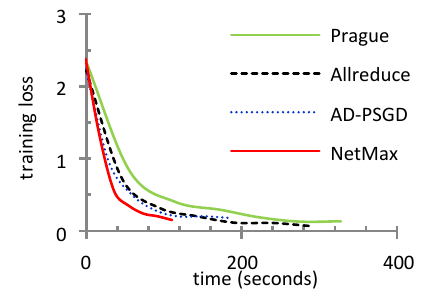} }  
		\vspace{-2ex} 
	\end{minipage} 
	\caption{Training MobileNet on MNIST.}
	\label{fig:loss-mnist}
	\vspace{-1ex}
\end{figure}

In this section, we present additional results for training the models on non-uniform data partitioning. We use the setup described in \autoref{subsec:nonuniform-data}. When training ResNet18 on CIFAR10, \ourtech{} achieves almost the same convergence rate compared to the other competitors due to the simplicity of classifying 10 classes, as shown in \autoref{fig:loss-cifar10}. When training ResNet18 on Tiny-ImageNet, \ourtech{} achieves slightly lower convergence rate while converging much faster over time than the other competitors, as shown in \autoref{fig:loss-tiny}. The final test accuracy of ResNet18 on Tiny-ImageNet for all the approaches is approximately 57\%, as shown in \autoref{tab:acc-nonuniform-data}. Training models on Tiny-ImageNet can hardly achieve a high accuracy since Tiny-ImageNet is only a subset of ImageNet and it is hard to train a model with high accuracy on such a complex dataset without enough data samples.

When training MobileNet on MNIST with non-IID data distribution, \ourtech{}~achieves a lower convergence rate compared to the other competitors, as shown in Fig.~\ref{fig:loss-mnist:a}. However, the lost convergence rate does not surpass the improvement in converging time, as shown in Fig.~\ref{fig:loss-mnist:b}. \ourtech{} achieves $2.45\times$, $2.35\times$, and $1.39\times$ speedup in terms of converging time compared to Prague, Allreduce-SGD, and AD-PSGD, respectively. Moreover, the improvement in converging time achieved by \ourtech{} can be enlarged when the model is larger or the network congestion is more serious, as shown in \autoref{fig:loss-imagenet} and \autoref{fig:loss-cifar100}.

\subsection{Distributed Training Across Clouds} 
\label{apendex:f}
We conducted experiments on a public cloud by training models across six Amazon EC2 regions, as summarized in Table~\ref{tab:lost-labels-clouds}. In each EC2 region, we created a \textit{c5.4$\times$large} instance which has 16 vCPU and $32$GB memory, running Ubuntu 16.04 LTS.

We train two CNN models (MobileNet and GoogLeNet) on the MNIST dataset to evaluate the performance of our proposed \ourtech{} approach. The parameter numbers of MobileNet and GoogLeNet are approximately 4.2M and 6.8M, respectively. Since users in different regions may have different habits and, thus, different data distribution, we sort the training dataset by labels, and assign part of the labels to the worker nodes, as described in Table~\ref{tab:lost-labels-clouds}. In addition, each worker node has the complete test dataset.

\begin{table}[tp]
	\centering
	\caption{Data distribution across cloud regions}
	\begin{tabular}{l|l|l|l}
		\hline
		Regions&Lost labels&Regions&Lost labels\\
		\hline
		US West&0, 1, 2&Mumbai&4, 5, 6 \\
		US East&1, 2, 3&Singapore&5, 6, 7\\
		Ireland&2, 3, 4&Tokyo&6, 7, 8\\
		\hline
	\end{tabular}
	\label{tab:lost-labels-clouds}
	\vspace{-1ex}
\end{table}

\begin{figure}[!t]
	\centering 
	\begin{minipage}[b]{0.23\textwidth}
		\centering
		\subfigure[MobileNet] { \label{fig:acc-multi-clouds:a}     \includegraphics[width=1\textwidth]{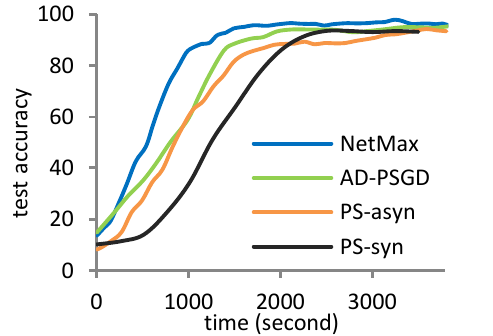} }
		\vspace{-2ex}
	\end{minipage}
	\hspace{0.01in}
	\begin{minipage}[b]{0.23\textwidth}
		\centering 
		\subfigure[GoogLeNet] { \label{fig:acc-multi-clouds:b}     \includegraphics[width=1\textwidth]{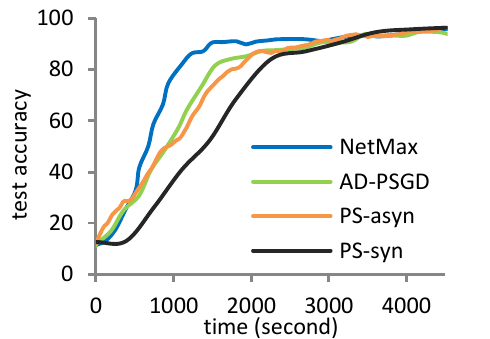} }  
		\vspace{-2ex} 
	\end{minipage} 
	\caption{Test accuracy varies with time. \ourtech{} converges much faster over time compared to other approaches.}
	\label{fig:acc-multi-clouds}
	\vspace{-1ex}
\end{figure}

The test accuracy of \ourtech{} converges much faster over time compared to the other approaches, as shown in \autoref{fig:acc-multi-clouds}. We observe that parameter server with synchronous training (PS-syn) is the slowest since its training speed is determined by the slowest link connecting the worker nodes and the parameter server. Parameter server with asynchronous training (PS-asyn) significantly outperforms PS-syn on model training, but it is slightly slower than AD-PSGD, the state-of-the-art asynchronous decentralized PSGD. On the other hand, AD-PSGD is much slower than \ourtech{}. In summary, \ourtech{} converges 1.9$\times$, 1.9$\times$, and 2.1$\times$ faster than AD-PSGD, PS-asyn and PS-syn, respectively.
}
\\
}
{
}

\end{document}